\documentclass[superscriptaddress,notitlepage,nofootinbib,a4paper]{revtex4-1}
\usepackage[UKenglish]{babel}
\usepackage{dsfont}
\usepackage{amssymb}
\usepackage{mathrsfs}
\usepackage{amsmath}
\usepackage{amsthm}
\usepackage{amsfonts}
\usepackage{mathtools}
\usepackage{footnote}
\usepackage{graphicx}
\usepackage{verbatim}
\usepackage{bbm}
\usepackage{tikz}
\usepackage{pgfplots}
\usepackage[left=3.3cm,right=3.3cm]{geometry}
\usepackage[hyperfootnotes=false]{hyperref}
\usepackage{natbib}
\usepackage{fancyhdr} 
\usepackage{gensymb}
\fancypagestyle{newstyle}{
\fancyhf{} 
\fancyfoot[R]{\thepage\centering} 

}
\usepgfplotslibrary{fillbetween}
\usetikzlibrary{patterns}

\numberwithin{equation}{section}

\newtheorem{thm}{Theorem}[section]
\newtheorem{lem}[thm]{Lemma}

\newtheorem{cor}[thm]{Corollary}
\newtheorem{prop}[thm]{Proposition}

\newtheorem*{lem*}{Lemma}

\newcommand\ve{\varepsilon}
\newcommand\vf{\varphi}
\newcommand\nn{\nonumber}

\newcommand{\JB}{J_{\mathrm{b}}}

\newcommand{\IIm}{[-\ve^{-{1\over 2}},\ve^{-{1\over 2}}]}
\newcommand{\IIp}{[-\ve^{-1},\ve^{-1}]}

\newcommand{\vei}{\ve^{-1}}
\newcommand{\veim}{\ve^{-{1\over 2}}}
\newcommand{\de}{\mathrm{d}}
\newcommand{\atanh}{\mathrm{arc}\tanh}
\newcommand{\sqve}{\sqrt{\ve}}
\newcommand{\mk}{m^{(\kappa)}}
\newcommand{\chib}{\chi_{\beta}}
\newcommand{\mmac}{\hat{m}_{\ve}^{\mathrm{macro}}}

\pgfplotsset{compat=1.12}
\pgfplotsset{grid style={gray!40}}

\begin{document}

\title{Current with ``wrong'' sign and phase transitions}
\author{Roberto Boccagna}
\affiliation{Gran Sasso Science Institute, Viale F. Crispi 7, 67100 L'Aquila, Italy}
\begin{abstract}
\noindent
We prove that under certain conditions, phase separation is enough to sustain a regime in which current flows along the concentration gradient, a phenomenon which is known in the literature as \textit{uphill diffusion}. The model we consider here is a version of that proposed in \cite{GIAC}, which is the continuous mesoscopic limit of a $1d$ discrete Ising chain with a Kac potential. The magnetization profile lies in the interval $\IIp$, $\ve>0$, staying in contact at the boundaries with infinite reservoirs of fixed magnetization $\pm\mu$, $\mu\in(m^*\left(\beta\right),1)$, where $m^*\left(\beta\right)=\sqrt{1-1/\beta}$, $\beta>1$ representing the inverse temperature. At last, an external field of Heaviside-type of intensity $\kappa>0$ is introduced. According to the axiomatic non-equilibrium theory, we derive from the mesoscopic free energy functional the corresponding stationary equation and prove the existence of a solution, which is antisymmetric with respect to the origin and discontinuous in $x=0$, provided $\ve$ is small enough. When $\mu$ is metastable, the current is positive and bounded from below by a positive constant independent of $\kappa$, this meaning that both phase transition as well as external field contributes to uphill diffusion, which is a regime that actually survives when the external bias is removed. 
\end{abstract}

\maketitle

\section{Introduction}

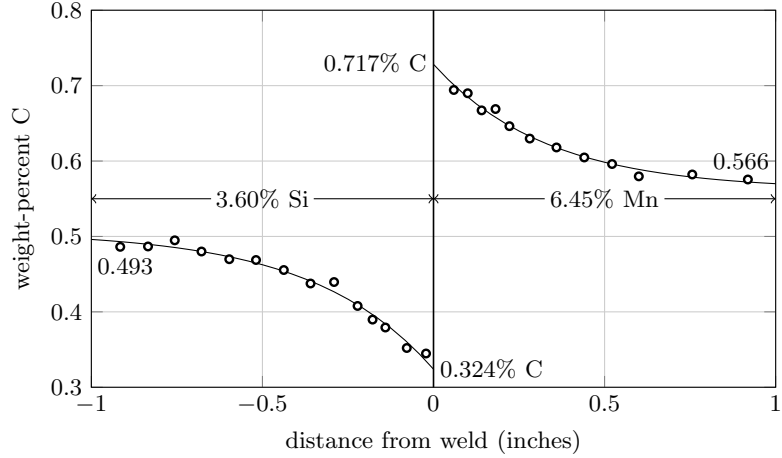
\begin{figure}[htbp!]
\centering
\begin{tikzpicture}[trim axis left, trim axis right]
\begin{axis}[
grid=both,
width=9cm,
height=5cm,
scale only axis,
ylabel=$\text{weight-percent C}$,
xlabel=$\text{distance from weld (inches)}$,
ymin=0.30,
ymax=0.80,
xmin=-1,
xmax=1,
xtick={-1.0,-0.5,0,0.5,1.0},
ytick={0.3,0.4,0.5,0.6,0.7,0.8},
]
\addplot[only marks,mark=*,mark size=0.5mm,mark options={color=black,line width=0.9pt,fill=white},color=black]
table{SiMn.dat};
\addplot[domain=0.001:1]{0.561987+0.166301*exp(-3.03827*x)};
\addplot[domain=-1:0.001]{0.506756-0.182751*exp(2.83615*x)};
\draw[<-,thin] (axis cs:-1,0.55) -- (axis cs:-0.65,0.55);
\draw[->,thin] (axis cs:-0.35,0.55) -- (axis cs:0,0.55);
\draw[<-,thin] (axis cs:0,0.55) -- (axis cs:0.325,0.55);
\draw[->,thin] (axis cs:0.675,0.55) -- (axis cs:1,0.55);
\draw[black,semithick] (axis cs:0,0.3) -- (axis cs:0,0.8);
\node[] at (axis cs: -0.5,0.55) {3.60$\%$ Si};
\node[] at (axis cs: 0.5,0.55) {6.45$\%$ Mn};
\node[] at (axis cs: -0.17,0.73) {0.717$\%$ C};
\node[] at (axis cs: 0.17,0.324) {0.324$\%$ C};
\node[] at (axis cs: -0.9,0.46) {0.493};
\node[] at (axis cs: 0.9,0.6) {0.566};
\end{axis}
\end{tikzpicture}
\caption{Uphill diffusion of carbon in the Fe-Si-Mn-C system (figure extrapolated from \cite{LSD}). Carbon diffuses from an austenite of carbon content of about 0.32 wt. $\%$ to an austenite of carbon content of about 0.72 wt. $\%$.}\label{fig1}
\end{figure}

\noindent
The issue in this paper concerns the relation between diffusion and concentration in the Fick's Law setting. As already noticed by W. Nernst \cite{NERNST} and L. Onsager \cite{ONS}, it may happen that current flows along the direction of the density gradient, a phenomenon which is known as \textit{uphill diffusion}. L. S. Darken was among the first who provided experimental evidences for diffusion of a given element towards a region of higher concentration in systems with more than two components, being also able to reduce it to a physically coherent framework \cite{DARKEN1,DARKEN2,LSD}. The experimental setup was like that: pairs of steel containing a slightly difference in carbon content, but a remarkable difference in the alloy content, were welded together and held in a furnace at a temperature of about $1050^{\circ}\mathrm{C}$ to let carbon diffuse. Several specimen were prepared, differing in the two doping substances, mostly silicon, manganese and molybdenum. Figure \ref{fig1} shows carbon distribution in a Fe-Si-Mn compound after $10$ days. Despite the initial concentration of carbon at left and right edges were about $0.493$ and $0.566$ wt. $\%$ respectively, carbon diffuses from the left to the right in the Fe-Si alloy and from the right to the left in the Fe-Mn alloy. As pointed out by the author, the explanation for that lies in the fact that silicon decreases the chemical affinity of carbon, while manganese increases it. This results in a driving force acting in the opposite direction with respect to the concentration gradient. This mechanism works until also dopants diffuse towards the steels; nevertheless, this happens in very large times (compared with the characteristic diffusion times of carbon), so that uphill regime reduces to a relatively long transient before standard diffusion establishes again. However, when the dopant rates in the steels are reduced, the difference in carbon concentration between the edges and the welding point gets smaller. If the doping substances are completely removed, the distribution becomes flat in each steel and this regime breaks off. Nowadays, this phenomenon is widely exploited by industries in the processes which involve purification of metals, and it is even more general than described so far (see \cite{KR1,KR2} for a nice overview on uphill diffusion in multicomponent systems). Indeed, the purpose of this work is to show that uphill diffusion may persist also in one-component systems when a phase transition occurs. 
\newline
The setup we have in mind is to constrain the system in a finite region which is in contact with reservoirs as in the Fourier-Fick usual setup. Our model is a non-local version of the Ginzburg-Landau free energy functional with an additional term which comes from a piecewise constant external field. In such model, magnetization plays the role of carbon in the Darken's experiment, while the external field plays the role of the doping substances. As we shall see, when a phase transition occurs there is a current which flows along the concentration gradient, provided the intensity of the field and the fixed magnetization of the reservoirs are chosen in a feasible way. Moreover, we will prove that in this case uphill diffusion regime persists also when the external field is removed.

\section{Model, Background, Main Results}

\subsection{Axiomatic non equilibrium theory}

\subsubsection{Mesoscopic equation with external magnetic field}

\noindent
%We recommend \cite{DMPT} and \cite{EP} for a detailed treatment of mesoscopic scaling. 
%\newline
For notational convenience, indicate:
\begin{equation}
\Lambda\coloneqq\IIp,\qquad \Lambda^c\coloneqq\mathbb{R}\setminus\IIp
\end{equation}
and let $m_{\Lambda}\in L^{\infty}\left(\Lambda,\left[-1,1\right]\right)$, $m_{\Lambda^c}\in L^{\infty}\left(\Lambda^c,\left[-1,1\right]\right)$, $m_{\Lambda}$ being the magnetization density of the bulk and $m_{\Lambda^c}$ the magnetization of the reservoirs, namely $m_{\Lambda^c}\left(x\right)=-\mu$ for $x\in\left(-\infty,-\vei\right)$ and $m_{\Lambda^c}\left(x\right)=\mu$ for $x\in\left(\vei,\infty\right)$, $\mu\in\left(0,1\right)$. Let $h^{\mathrm{ext}}\in L^{\infty}\left(\Lambda\right)$ the external magnetic field. In the sequel, $h^{\mathrm{ext}}\left(x\right)\coloneqq \kappa\,\mathrm{sign}\left(x\right)$, $\kappa>0$, for any $x\in\mathbb{R}$. Consider then the mesoscopic Ginzburg-Landau free energy functional:
\begin{eqnarray}
&&\mathcal{F}_{\beta}\left[m_{\Lambda}\mid m_{\Lambda^c},h^{\mathrm{ext}}\right] = \mathcal{F}_{\beta}\left[m_{\Lambda}\mid h^{\mathrm{ext}}\right] - \int_{\Lambda}\int_{\Lambda^c}J\left(x,y\right)m_{\Lambda}\left(x\right)m_{\Lambda^c}\left(y\right)\de x\,\de y, \label{freeen}\\
&&\mathcal{F}_{\beta}\left[m_{\Lambda}\mid h^{\mathrm{ext}}\right] = -{1\over\beta}\int_{\Lambda}S\left(m_{\Lambda}\left(x\right)\right)\de x -\int_{\Lambda}h^{\mathrm{ext}}\left(x\right)m_{\Lambda}\left(x\right)\de x\nn\\
&&\hspace{2.675cm}- {1\over 2} \int_{\Lambda}\int_{\Lambda}J\left(x,y\right)m_{\Lambda}\left(x\right)m_{\Lambda}\left(y\right)\de x\,\de y
\end{eqnarray}
where $S\left(m\right)$ is the standard binary entropy for an Ising spin system:
\begin{equation}\label{entr}
S\left(m\right)=-{1+m\over 2}\log\left({1+m\over 2}\right)-{1-m\over 2}\log\left({1-m\over 2}\right)
\end{equation}
and $J\left(x,y\right)$, $x,y\in\mathbb{R}$ is a transition probability kernel with properties:
\begin{itemize}
\item[-] $J\left(x,y\right)=J\left(0,\left|x-y\right|\right)$
\item[-] $J\left(0,x\right)\in C_c^{\infty}\left(\left[-1,1\right]\right)$, $\int J\left(\,\cdot\,,x\right)\de x=1$
\item[-] $\sup_{x\ge 0}J\left(0,x\right)=1$
\item[-] $J\left(0,x\right)$ is strictly decreasing in $x\in\left[0,1\right]$.
\end{itemize}
We suppose the magnetization to evolve in time according to a gradient dynamics for any $t\ge 0$:
\begin{equation}\label{prim}
{\partial m_{\Lambda}\over \partial t }\left(x,t\right)=-{\partial I\over \partial x}\left(x,t\right), \qquad
I\left(x,t\right)\coloneqq-\chi_{\beta}\left(m_{\Lambda}\left(x,t\right)\right) {\de\over\de x} {\delta\over\delta m_{\Lambda}\left(x,t\right)}\mathcal{F}_{\beta}\left[m_{\Lambda}\mid m_{\Lambda^c},h^{\mathrm{ext}}\right] 
\end{equation}
where $I$ represents the current and $\delta$ denotes functional derivative. $\chi_{\beta}\left(m\right)$ is the magnetic susceptibility which is set equal to:
\begin{equation}
\chi_{\beta}\left(m\right)\coloneqq\beta\left(1-m^2\right).
\end{equation}
We look for a stationary solution of \eqref{prim}, that is a couple $m\in L^{\infty}\left(\IIp\right)$, $I\in\mathbb{R}$ such that:
\begin{equation}\label{frect}
{\de m\over\de t}\left(x,t\right)=0, \qquad I\left(x,t\right)\equiv I=\mathrm{const}.
\end{equation}
In the free boundary case with $\kappa=0$, it has been established by E. Presutti et al. \cite{EP1,EP2} that problem \eqref{frect} admits a one-parameter family of solutions $m\in C^{\infty}\left(\mathbb{R},\left[-1,1\right]\right)$ with $I\equiv 0$, for any $\beta>1$. These profiles, called \textit{instantons}, are monotone functions connecting the two pure phases $\pm m_{\beta}$, $m_{\beta}$ the positive solution of the mean-field equation $m_{\beta}=\tanh\left(\beta m_{\beta}\right)$ at $\beta>1$. Furthermore, they are limit orbits of a gradient-type dynamics, thus truly minimizers of the corresponding free energy functional. We recommend \cite{EP} for an exhaustive treatment of that.
\newline
By symmetry, we expect in our case a solution of \eqref{frect}, if any, to be an odd function, its shape depending on the values of $\kappa$ and $\mu$. Sketchily, on the positive half line, the magnetization profile should connect $m_{\beta,\kappa}$ to the boundary value $\mu$, where $m_{\beta,\kappa}$ is the positive solution of $m_{\beta,\kappa}=\tanh\left[\beta\left(m_{\beta,\kappa}+\kappa\right)\right]$. In fact, $m_{\beta,\kappa}$ is an equilibrium value in the mean-field model. At fixed $\beta>1$, $m_{\beta,\kappa}>m_{\beta}$ so that we expect the current to be positive if $\mu<m_{\beta}$, i.e. when $\mu$ lies in the spinodal region $\left(0,m_{\beta}\right)$. Indeed, we will prove that a stationary solution with positive current does exist at any $\kappa>0$ provided $\mu\in\left(m^*\left(\beta\right),m_{\beta}\right)$, where $m^*\left(\beta\right)=\sqrt{1-1/\beta}$, and that uphill diffusion regime survives also in the limit $\kappa\downarrow 0$, since the current is bounded from below by a positive constant which is independent of $\kappa$. Of course, stationary solutions do exist also when $\mu>m_{\beta}$, although the sign of the current is negative when $\mu>m_{\beta,\kappa}$. The method we are going to exploit to prove this result does not guarantee the solution at fixed $\kappa$ and $\mu$ to be unique. However, numerical simulations suggest that this should be the case at least when $\mu$ is ``close'' to $m_{\beta}$. 
\newline
It is worth remarking here that the problem we address has its own interest in the attempt of establishing a well posed mathematical theory for phase transitions in the non-equilibrium setting, which is still lacking despite much has been done \cite{JL1,JL2,JL3}. In particular, our work goes towards the direction of proving existence of instantons for open systems. In fact, the guess is that antisymmetric profiles survive when the limit $\kappa\downarrow 0$ is performed. This claim is not a straight consequence of the results presented here, and a work dedicated to this proof of this will follow; nevertheless, such result turns out to be not so interesting from a physical point of view, since numerical simulations indicate that instantons are energetically unstable and never observed in practice \cite{CDMP,CGGV}.
\newline
Recalling \eqref{prim}, we get for any $x\in\Lambda$ (dependence on time is neglected below in the notation):
\begin{equation}
{\delta\mathcal{F}_{\beta}\left[m_{\Lambda}\mid h^{\mathrm{ext}}\right]\over\delta m_{\Lambda}\left(x\right)}={1\over\beta}\,\atanh\left(m_{\Lambda}\left(x\right)\right)-h^{\mathrm{ext}}\left(x\right)-\int_{\Lambda}J\left(x,y\right)m_{\Lambda}\left(y\right)\de y.
\end{equation}
Considering that
\begin{equation}
{\delta\over\delta m_{\Lambda}\left(x\right)}\left(\int_{\Lambda}\int_{\Lambda^c}J\left(x,y\right)m_{\Lambda}\left(x\right)m_{\Lambda^c}\left(y\right)\de x\,\de y\right)=\int_{\Lambda^c}J\left(x,y\right)m_{\Lambda^c}\left(y\right)\de y,
\end{equation}
we obtain
\begin{equation}
{\delta\mathcal{F}_{\beta}\left[m_{\Lambda}\mid m_{\Lambda^c},h^{\mathrm{ext}}\right]\over\delta m_{\Lambda}\left(x\right)}={1\over\beta}\,\atanh\left(m_{\Lambda}\left(x\right)\right)-h^{\mathrm{ext}}\left(x\right)-\left(J\ast m\right)\left(x\right)
\end{equation}
where, by definition, $m\left(x\right)=m_{\Lambda}\left(x\right)$ if $x\in\Lambda$ and $m\left(x\right)=m_{\Lambda^c}\left(x\right)$ if $x\in\Lambda^c$. By the choice of $m_{\Lambda_c}$, it is worth writing, for any $x\in\Lambda$:
\begin{eqnarray}
\left(J\ast m\right)\left(x\right)&=&\int_{\Lambda}J\left(x,y\right)m_{\Lambda}\left(y\right)\de y-\mu\int_{-\infty}^{-\vei}J\left(x,y\right)\de y +\mu\int_{\vei}^{+\infty}J\left(x,y\right)\de y\nn\\
&=&\int_{\Lambda}J\left(x,y\right)m_{\Lambda}\left(y\right)\de y-\int_{-\infty}^{-\vei}J\left(x,y\right)\left(\int m_{\Lambda}\left(z\right)\delta\left(z+\vei\right)\de z\right)\de y\nn\\
&+&\int_{\vei}^{+\infty}J\left(x,y\right)\left(\int m_{\Lambda}\left(z\right)\delta\left(z-\vei\right)\de z\right)\de y
\end{eqnarray}
so that, if we define in the sense of distributions
\begin{equation}
\JB\left(x,y\right)\stackrel{\mathcal{D}'}{\coloneqq}
J\left(x,y\right)\mathbf{1}_{\left|y\right|<\ve^{-1}}+{1\over 2}b_{\ve}\left(\left|x\right|\right)\left[\delta\left(\ve^{-1}-y\right)+\delta\left(\ve^{-1}+y\right)\right]\mathbf{1}_{\left|y\right|\ge\vei},
\end{equation}
where
\begin{equation}
b_{\ve}\left(x\right)\coloneqq\int_{\ve^{-1}}^{+\infty}J\left(x,y\right)\de y,
\end{equation}
we get:
\begin{equation}
\left(J\ast m\right)\left(x\right)=\int_{-\infty}^{+\infty}\JB\left(x,y\right)m_{\Lambda}\left(x\right)\de y =\left(\JB\star m_{\Lambda}\right)\left(x\right)=\left(\JB\star m\right)\left(x\right).
\end{equation}
In this redefinition of the convolution kernel, we restricted the problem to the interval $\IIp$, avoiding the region $\Lambda^c$. From the properties of $J\left(x,y\right)$, $\mathrm{supp}\left(b_{\ve}\right)=[\ve^{-1}-1,\ve^{-1}]$, and that $\int\JB\left(x,y\right)\de y=1$ for any $x\in\Lambda$. In this way (dropping the suffix $\Lambda$):
\begin{equation}
{\delta\mathcal{F}_{\beta}\left[m_{\Lambda}\mid m_{\Lambda^c},h^{\mathrm{ext}}\right]\over\delta m_{\Lambda}\left(x\right)}={1\over\beta}\,\atanh\left(m\left(x\right)\right)-h^{\mathrm{ext}}\left(x\right)-\left(\JB\star m\right)\left(x\right).
\end{equation}
Following a strategy established in \cite{DMPT,DMOPc} and also used in \cite{RB}, we change variable by setting
\begin{equation}\label{hdef}
\widetilde{h}\left(x\right)\coloneqq{1\over\beta}\,\atanh\left(m\left(x\right)\right)-h^{\mathrm{ext}}\left(x\right)-\left(\JB\star m\right)\left(x\right),
\end{equation}
so that
% PERò NEL NOSTRO CASO NON è DERIVABILE
\begin{equation}\label{idef}
I\left(x\right) = -\chi_{\beta}\left(m\left(x\right)\right){\de \widetilde{h}\over \de x}\left(x\right).
\end{equation}
Then, inverting \eqref{hdef} and integrating \eqref{idef} we end up with system:
\begin{equation}\label{sys}
\begin{dcases}
m\left(x\right)=\tanh\Big\{\beta\Big[\left(\JB\star m\right)\left(x\right)+h^{\mathrm{ext}}\left(x\right)+\widetilde{h}\left(x\right)\Big]\Big\} \\
\widetilde{h}\left(x\right)=-\widetilde{h}\left(x_0\right)-I\int_{x_0}^{x}\chi_{\beta}^{-1}\left(m\left(y\right)\right)\de y
\end{dcases}
\qquad x\in \IIp.
\end{equation}
We change variable again by defining $h\left(x\right)\coloneqq h^{\mathrm{ext}}\left(x\right)+\widetilde{h}\left(x\right)$. We are looking for antisymmetric solutions in $\IIp$, then we set equal to $0$ the constant of integration in the second of \eqref{sys}. We also expect the current to be of order $\ve$ (see \cite{RB}), then we set $I\equiv j^{(\ve,\kappa)}\ve$, where $j^{(\ve,\kappa)}=O\left(1\right)$. Thus \eqref{sys} becomes
\begin{equation}\label{syst}
\begin{dcases}
m^{(\ve,\kappa)}\left(x\right)=\tanh\Big\{\beta\Big[\big(\JB\star m^{(\ve,\kappa)}\big)\left(x\right)+h^{(\ve,\kappa)}\left(x\right)\Big]\Big\} \\
h^{(\ve,\kappa)}\left(x\right)=\kappa\,\mathrm{sign}\left(x\right)-j^{(\ve,\kappa)}\ve\int_{x_0}^{x}\chi_{\beta}^{-1}\left(m^{(\ve,\kappa)}\left(y\right)\right)\de y
\end{dcases}
\qquad x\in\IIp
\end{equation}
which is the problem we deal with, given $m^{(\ve,\kappa)}\,(-\vei)=-\mu$, $m^{(\ve,\kappa)}\,(\vei)=\mu$.

\begin{comment}
\newline
\newline
The paper is organized as follows:
\newline
- Section 2: we recall axiomatic non-equilibrium theory and introduce the model. Stationary profiles are obtained as critical points of the corresponding mesoscopic functional. Main theorems are stated, together with propositions we are going to use to prove existence of an antisymmetric solution; an iterative strategy is defined.
\newline
- Section 3: we prove existence of instantons at $\kappa>0$ in the infinite volume case.
\newline
- Section 4: we provide estimates for the main operator carrying the interaction, which guarantees its invertibility.
\newline
- Section 5: we enter the iterative scheme and prove convergence of the first iteration.
\newline
- Section 6: we prove that iterations as defined in Section 2 converge to a solution of the stationary problem for any $\kappa>0$. 
\newline
- Section 7: we prove existence of solutions for any choice of opposite boundary conditions lying in the stable or metastable region.
\end{comment}

\subsection{Existence}

\noindent
Let $\mathscr{M}_{\beta}\coloneqq\left\{m\in\mathbb{R} \mid m^*\left(\beta\right)<m<1\right\}$, $\beta>1$. Our main results are:
\begin{thm}\label{t1}
For any $\kappa>0$ there is an increasing, antisymmetric function $m^{(\kappa)}$ such that
\begin{equation}\label{free}
m^{(\kappa)}\left(x\right)=\tanh\Big\{\beta\Big[\big(J\ast m^{(\kappa)}\big)\left(x\right)+\kappa\,\mathrm{sign}\left(x\right)\Big]\Big\}, \qquad x\in\mathbb{R}
\end{equation}
Moreover, $m^{(\kappa)}$ has limits:
\begin{equation}
\lim_{x\to 0} m^{(\kappa)}\left(x\right) = \tanh\left(\beta\kappa\right)
\end{equation}
and
\begin{equation}
\lim_{x\to\infty} m^{(\kappa)}\left(x\right) = m_{\beta,\kappa}, \qquad m_{\beta,\kappa} = \tanh\big\{\beta\big[m_{\beta,\kappa}+\kappa\big]\big\}.
\end{equation}
\end{thm}
\noindent
Equation \eqref{free} is the analogous of \eqref{syst} in the infinite size setting. It can be proved in fact that \eqref{free} comes from a variational problem as well. As we shall see, $m^{(\kappa)}$ plays a crucial role in the construction of the solution to our problem.
\begin{thm}\label{t2}
For any $\kappa>0$ and $\mu\in\mathscr{M}_{\beta}$, there is $\ve_0>0$ such that for any $\ve<\ve_0$ problem \eqref{syst} admits a solution $\big(m^{(\ve,\kappa)},j^{(\ve,\kappa)}\big)$, with $m^{(\ve,\kappa)}$ antisymmetric    in $\IIm$ and satisfying $m(\vei)=\mu$. $m^{(\ve,\kappa)}$ is differentiable in $(0,\vei]$, and $j^{(\ve,\kappa)}$ is positive provided $\mu\in\left(m^*\left(\beta\right),m_{\beta}\right)$.
\end{thm}
\noindent
For the sake of notational simplicity, we do not emphasize anymore dependence on $\ve$ and $\kappa$ from now on. Since we expect the solution of \eqref{syst} to be antisymmetric, we restrict to the positive subset $(0,\vei]$, so that \eqref{syst} reads
\begin{equation}\label{systm}
\begin{dcases}
m\left(x\right)=\tanh\Big\{\beta\Big[\left(\JB\star m\right)\left(x\right)+h\left(x\right)\Big]\Big\} \\
h\left(x\right)=\kappa-j\ve\int_{0}^{x}\chi_{\beta}^{-1}\left(m\left(y\right)\right)\de y
\end{dcases}
\qquad x\in(0,\vei]
\end{equation}
with $m\,(\vei)=\mu$, $\mu\in\mathscr{M}_{\beta}$. It can be checked that this formulation is equivalent at all to \eqref{syst} provided $m\left(-x\right)=-m\left(x\right)$ and $h\left(-x\right)=-h\left(x\right)$ for any $x\in[-\vei,0)$.

\begin{comment}
What we need to check is that this restriction actually keeps the solution antisymmetric. We thus impose $m\left(x\right)=-m\left(-x\right)$ for $x\in[-\vei,0)$ and verify that this condition holds true. It turns out that this is the case if $h\left(x\right)=-h\left(-x\right)$. Indeed, compute for any $x\in[-\vei,0)$:
\begin{eqnarray}
\left(\JB\star m\right)\left(x\right)&=&\int_{-\infty}^{+\infty}\JB\left(x,y\right)m\left(y\right)\de y =\int_{-\infty}^{+\infty}\JB\left(-x,-y\right)m\left(y\right)\de y\nn\\
&=&-\int_{+\infty}^{-\infty}\JB\left(-x,z\right)m\left(-z\right)\de z=\int_{-\infty}^{+\infty}\JB\left(-x,z\right)m\left(-z\right)\de z\nn\\
&=&-\int_{-\infty}^{+\infty}\JB\left(-x,z\right)m\left(z\right)\de z=-\left(\JB\star m\right)\left(-x\right).
\end{eqnarray}
In this positions
\begin{eqnarray}
m\left(x\right)=-m\left(-x\right)&=&-\tanh\Big\{\beta\Big[\left(\JB\star m\right)\left(-x\right)+
h\left(-x\right)\Big]\Big\}\nn\\
&=&\tanh\Big\{\beta\Big[-\left(\JB\star m\right)\left(-x\right)-
h\left(-x\right)\Big]\Big\}\nn\\
&=&\tanh\Big\{\beta\Big[\left(\JB\star m\right)\left(x\right)+
h\left(x\right)\Big]\Big\}\nn\\
\end{eqnarray}
thus, correspondence is preserved provided:
\begin{equation}
\begin{dcases}
m\left(x\right)=\tanh\Big\{\beta\Big[\left(\JB\star m\right)\left(x\right)+
h\left(x\right)\Big]\Big\}\\
h\left(x\right)=\kappa-j\ve\int_0^x\chi_{\beta}^{-1}\left(m\left(y\right)\right)\de y\\
m\,(\vei)=\mu\\
m\left(-x\right)=-m\left(x\right)\\
h\left(-x\right)=-h\left(x\right)
\end{dcases}
\end{equation}
for any $x\in(0,\vei]$.

\end{comment}

\subsubsection{Outline of the Proof}

\noindent
In the first part of the work, we prove that at fixed $j$ \eqref{systm} admits a continuous solution $\left(m,h\right)$ which satisfies a certain boundary condition $m\,(\vei)=\nu$. The proof works by iteration: we choose a starting pair $\left(m_0,h_0\right)$ and define a sequence of functions $\left(m_n,h_n\right)_{n=0}^{\infty}$ that converges to a pair $\left(m,h\right)$ which in turn solves \eqref{systm}. We do not know a priori the value of $\nu$ we end up with, so that we cannot conclude that \eqref{systm} has a solution for any fixed boundary condition in $\mathscr{M}_{\beta}$. Thus, we later prove that $j$ can be conveniently chosen in order to cover the whole region $\mathscr{M}_{\beta}$. The choice of $m_0$ is essential in this scheme; the starting profile has to be in fact ``close'' to what we suppose to be the solution of \eqref{systm}. Formally, in the nearby of $x=0$ we expect $m$ to be similar to the solution of the free boundary problem \eqref{free} if $\vei$ is large enough. As we will prove, $m^{(\kappa)}\left(x\right)$ converges exponentially fast to $m_{\beta,\kappa}$ for $x\to\infty$; thus, whatever the value of $\kappa$, we can always take $\ve^{-1}$ so large that there is $x'\in(0,\vei]$ so that $m^{(\kappa)}\left(x'\right)>m_{\beta}$. Conversely, in the region $(x',\vei]$, we expect $m^{(\kappa)}\left(x'\right)\sim m_{\beta,\kappa}$ to be connected to the boundary magnetization $\mu$ through a monotone profile. Since we want also $x'\gg 1$ to perform a suitable approximation further on, we fix $x'\equiv\veim$. Call $\mu_0\coloneqq m^{(\kappa)}\,(\veim)$; after the change of variable $r\equiv {\ve x-\sqve\over 1-\sqve}$, $r\in\left[0,1\right]$ for any $x\in(\veim,\vei]$, we consider the (macroscopic) free energy funcional: 
\begin{equation}\label{fmacro}
\mathcal{F}^{\mathrm{macro}}_{\beta;\mu_0,\mu}\left[m;h^{\mathrm{ext}}\right]=-{1\over\beta}\int_0^1 S\left(m\left(r\right)\right)\de r -{1\over 2}\int_0^1 m^2\left(r\right)\de r - \kappa m\left(x\right)
\end{equation}
which is obtained from \eqref{freeen} after taking the ``local'' limit $\ve\downarrow 0$ in which $J$ converges weakly to a Dirac delta. Functional \eqref{fmacro} is actually well defined since we restricted to the plus phase in the half line $x>0$ and, by the choice of the boundary condition, $m\left(r\right)\in\mathscr{M}_{\beta}$ for any $r\in\left[0,1\right]$. Again, according to the axiomatic theory, the (macroscopic) current flowing is given by
\begin{equation}\label{boh}
j^{\mathrm{macro}}=\chi_{\beta}\left(m\left(r\right)\right){\de\over\de r}{\delta\mathcal{F}^{\mathrm{macro}}_{\beta;\mu_0,\mu}\left[m;h^{\mathrm{ext}}\right]\over\delta m\left(r\right)}.
\end{equation}
Then, by \eqref{fmacro} and \eqref{entr}:
\begin{equation}
{\delta\mathcal{F}^{\mathrm{macro}}_{\beta;\mu_0,\mu}\left[m;h^{\mathrm{ext}}\right]\over\delta m\left(r\right)}=-{1\over\beta}S'\left(m\left(r\right)\right)-m\left(r\right)
\end{equation}
with $S'\left(m\right)=-\atanh\left(m\right)$, so \eqref{boh} becomes
\begin{equation}
j^{\mathrm{macro}}=-\left[1-\chi_{\beta}\left(m\left(r\right)\right)\right]{\de m\over \de r}\left(r\right).
\end{equation}
Therefore, a profile $m^{\mathrm{macro}}$ which makes stationary the macroscopic functional \eqref{fmacro} solves the ODE:
\begin{equation}\label{mmacro}
\begin{dcases}
{\de m^{\mathrm{macro}}\over \de r}\left(r\right)=-j^{\mathrm{macro}}\left[1-\chi_{\beta}\left(m^{\mathrm{macro}}\left(r\right)\right)\right]^{-1},\qquad r\in\left[0,1\right]\\
m^{\mathrm{macro}}\left(0\right)=\mu_0, \;\;\; m^{\mathrm{macro}}\left(1\right)=\mu
\end{dcases}
\end{equation}
where $j^{\mathrm{macro}}$ is fixed by $\mu_0$ and $\mu$. Integrating \eqref{mmacro} from $0$ to $r\in\left[0,1\right]$, we get
\begin{equation}\label{jr}
j^{\mathrm{macro}}r=\left(\beta-1\right)\left[m^{\mathrm{macro}}\left(r\right)-\mu_0\right]-{\beta\over 3}\left[(m^{\mathrm{macro}})^3\left(r\right)-\mu_0^3\right],
\end{equation}
with
\begin{equation}
j^{\mathrm{macro}}=\left(\beta-1\right)\left(\mu-\mu_0\right)-{\beta\over 3}\left(\mu^3-\mu_0^3\right).
\end{equation}
Observe that in \eqref{jr} $r$ is a smooth function of $m$ and then invertible because $\mathrm{sign}\,{\de m^{\mathrm{macro}}\over \de r}=\mathrm{sign}\,j^{\mathrm{macro}}=\mathrm{const}$. Moreover, $m^{\mathrm{macro}}\in C^{\infty}\left(\left[0,1\right],\left[-1,1\right]\right)$. Back to the mesoscopic units, we set then $m_0$ to be the piecewise function
\begin{equation}
m_0\left(x\right)\coloneqq
\begin{dcases}\label{m0}
m^{(\kappa)}\left(x\right)\qquad &x\in(0,\veim]\\
m^{\text{macro}}\left(\frac{\ve x-\sqve}{1-\sqve}\right)\eqqcolon \hat{m}_{\ve}^{\mathrm{macro}}\left(x\right)\qquad &x\in(\veim,\vei],
\end{dcases}
\end{equation}
which we expect to differ of order $\sqve$ to the solution of \eqref{systm} for any $x\in(0,\vei]$. Notice that $m_0$ is continuous in $(0,\vei]$.
\newline
Define the map $\mathsf{T}^{(\kappa)}$ on $L^{\infty}\left((0,\vei]\right)$:
\begin{equation}
\mathsf{T}m\left(x\right)\coloneqq\kappa-j\ve\int_0^x\chi_{\beta}^{-1}\left(m\left(y\right)\right)\de y.
\end{equation}
The following proposition defines the first iterate.
\begin{prop}\label{p.3}
For any $\kappa>0$ there is $m_1\in C^1\left((0,\vei],[0,1]\right)$ which solves
\begin{equation}\label{m1}
m_1\left(x\right)=\tanh\Big\{\beta\Big[\left(\JB\star m_1\right)\left(x\right)+
h_0\left(x\right)\Big]\Big\}
\end{equation}
where $h_0=\mathsf{T}^{(\kappa)}m_0$, provided $\ve$ is small enough.
\end{prop}
\noindent
Subsequent iterations are explicitely defined through
\begin{prop}\label{p.4}
For any $\kappa>0$ and $n\in\mathbb{N}$ there is $m_{n+1}\in C^1\left((0,\vei],[0,1]\right)$ which solves
\begin{equation}\label{mn}
m_{n+1}\left(x\right)=\tanh\Big\{\beta\Big[\left(\JB\star m_{n+1}\right)\left(x\right)+
h_n\left(x\right)\Big]\Big\}
\end{equation}
where $h_n=\mathsf{T}^{(\kappa)}m_n$, $m_0$ as in \eqref{m0}, provided $\ve$ is small enough. Furthermore, there exists a pair $\left(m,h\right)$, $m\in C^1\left((0,\vei],[0,1]\right)$ such that
\begin{equation}
\lim_{n\to\infty}\left\|m_n-m\right\|_{\infty}=0, \qquad \lim_{n\to\infty}\left\|h_n-h\right\|_{\infty}=0,
\end{equation}
$\left(m,h\right)$ solving problem \eqref{systm} with $m\,(\vei)=\nu$, $\nu\in\mathscr{M}_{\beta}$.
\end{prop}
\noindent
Existence for any $\mu\in\mathscr{M}_{\beta}$ follows from:
\begin{prop}\label{p.5}
For any $\kappa>0$ and $\mu\in\mathscr{M}_{\beta}$ there is at least one $j\in\mathbb{R}$ such that $m\,(\vei;j)=\mu$.
\end{prop}
\noindent
Proposition \ref{p.5} closes the proof of Theorem \ref{t2}.

\section{Infinite size instanton at $\kappa>0$}

\subsection{Proof of Theorem \ref{t1}}

\subsubsection{Existence}

\noindent
Define the evolution semigroup $\mathsf{S}^{(\kappa)}_t$ on $L^{\infty}\left(\mathbb{R},\left[-1,1\right]\right)$ by setting $\mathsf{S}^{(\kappa)}_t m$ equal to the solution $m$ of the evolution equation:
\begin{equation}
{\mathrm{d}m\over\mathrm{d}t}\left(x,t\right) = -m\left(x,t\right) + \tanh\Big\{\beta\Big[\left(J\ast m\right)\left(x,t\right) +\kappa\,\mathrm{sign}\left(x\right)\Big]\Big\}, \qquad m\left(x,0\right)=m\left(x\right).
\end{equation}
We call $\mathsf{S}_t$ the semigroup when $\kappa=0$. Define $m_{\kappa,0}$ as the antisymmetric function given for $x\ge 0$ by:
\begin{equation}\label{in_cond}
m_{\kappa,0}\left(x\right) =
\begin{cases}
x m_{\beta,\kappa} \qquad &x\in\left[0,1\right]\\
m_{\beta,\kappa} \qquad &x\ge 1
\end{cases}
\end{equation}
then:
\begin{equation}\label{limits}
\lim_{t\to\infty} \mathsf{S}^{(\kappa)}_t m_{\kappa,0} = m^{(\kappa)}, \qquad \lim_{t\to\infty} \mathsf{S}^{(\kappa)}_t m_{0,0} = \bar{m}.
\end{equation}
Actually the proof of the existence of the limits is the key to the whole proof.
\begin{lem}
In the hypothesis $J\left(0,x\right)$ is a non increasing function of $x\ge 0$:
\begin{itemize}
\item[-] let $m$ and $\widetilde{m}$ be antisymmetric, non decreasing functions such that $\widetilde{m}\left(x\right)\ge  m\left(x\right)$ for all $x\ge 0$, then:
\begin{equation}
\mathsf{S}^{(\kappa)}_t \widetilde{m}\left(x\right) \ge \mathsf{S}^{(\kappa)}_t m\left(x\right)
\end{equation}
for all $x\ge 0$ and $t>0$.
\item[-] For all $x\ge 0$, $m^{(\kappa)}\left(x\right)>\bar{m}\left(x\right)$, more precisely:
\begin{equation}\label{claim2}
m^{(\kappa)}\left(x\right) \ge \bar{m}\left(x\right) + \beta\left(1-m^2_{\beta,\kappa}\right)\kappa.
\end{equation}
\end{itemize}
\end{lem}
\noindent
\textbf{Proof.} Let $T>0$ and $\Sigma$ the space of all bounded functions $\psi\left(x,t\right)$, $t\in\left[0,T\right]$, which for all $t$ are non decreasing, antisymmetric and with values in $\left[-1,1\right]$. Let $\mathsf{U}^{(\kappa)} : \Sigma \mapsto \Sigma$ be defined as:
\begin{equation}
\mathsf{U}^{(\kappa)}\psi\left(x,t\right) = \mathrm{e}^{-t}\psi\left(x,0\right) + \int_0^t\mathrm{e}^{-\left(t-s\right)}\tanh\Big\{\beta\Big[\left(J\ast\psi\left(x,s\right)\right)+\kappa\,\mathrm{sign}\left(x\right)\Big]\Big\}\,\de s.
\end{equation}
$\mathsf{U}\psi$ is defined as in \eqref{in_cond} with $\kappa=0$. Notice that if $\psi$ is a fixed point of $\mathsf{U}^{(\kappa)}$, then $\psi\left(x,t\right) = \left[\mathsf{S}^{(\kappa)}_t\left(\psi\left(\,\cdot\,,t\right)\right)\right]\left(x\right)$. We claim that if $\psi$, $\phi\in\Sigma$ and $\psi\left(x,t\right)\ge\phi\left(x,t\right)$ for all $x\ge 0$ and $t\in\left[0,T\right]$, then also $\mathsf{U}\psi\left(x,t\right)\ge\mathsf{U}^{(\kappa)}\phi\left(x,t\right)$ for all $x\ge 0$ and $t\in\left[0,T\right]$. Fix $t\in\left[0,T\right]$ and call $\theta\left(x\right)=\psi\left(x,t\right)-\phi\left(x,t\right)$, then the claim follows after proving that $\left(J\ast\theta\right)\left(x\right)\ge 0$ for all $x\ge 0$. This is so because for $x\ge 0$, by the antisymmetry of $\theta$:
\begin{equation}\label{J_theta}
\int J\left(x,y\right) \theta\left(y\right)\de y = \int_{y\ge 0}\left[J\left(x,y\right)-J\left(x,-y\right)\right]\theta\left(y\right)\de y
\end{equation}
and the right hand side is non negative because $J$ is non decreasing and $\theta\left(y\right)\ge 0$ for $y\ge 0$.
\newline
For $T>0$ small enough $\mathsf{U}^{(\kappa)}$ is a contraction in the sup norm in any subset of $\Sigma$ where the value of $\psi$ is fixed at $t=0$, same statement holding for $\mathsf{U}$. In any such subsets $\mathsf{U}^{(\kappa)}$ has a fixed point which is given by $\mathsf{S}^{(\kappa)}_t\left(\psi\left(x,0\right)\right)$, analogous statement holds for $\mathsf{U}$. By \eqref{J_theta} it then follows that $\mathsf{S}^{(\kappa)}_t\widetilde{m}\left(x\right)\ge\mathsf{S}_t m\left(x\right)$ for all $x\ge 0$ and all $t\in\left[0,T\right]$. The first claim in the theorem then follows by iteration over $T$.
\newline
We have proved so far that for all $x\ge 0$ and all $t\ge 0$:
\begin{equation}
\mathsf{S}^{(\kappa)}_t m_{\kappa,0}\ge \mathsf{S}_t m_{0,0}
\end{equation}
which proves by \eqref{limits} that $m^{(\kappa)}\left(x\right)\ge\bar{m}\left(x\right)$ for all $x\ge 0$. By \eqref{free}, after a Taylor expansion:
\begin{equation}
m^{(\kappa)}\left(x\right) - \bar{m}\left(x\right) = p^*\left(x\right)\left[\left(J\ast\big(m^{(\kappa)}-\bar{m}\big)\right)\left(x\right)+\kappa\,\mathrm{sign}\left(x\right)\right]
\end{equation}
where $p^*\left(x\right)$ is some value in the interval $\left[\beta\left[1-(m^{(\kappa)}\left(x\right))^2\right],\beta\left[1-(\bar{m}\left(x\right))^2\right]\right]$. As in \eqref{J_theta}, $\left(J\ast\left(m^{(\kappa)}-\bar{m}\right)\right)\left(x\right)\ge 0$ when $x\ge 0$ so that for such values of $x$
\begin{equation}
m^{(\kappa)}\left(x\right)-\bar{m}\left(x\right)\ge \beta\left[1-(m^{(\kappa)}\left(x\right))^2\right]\kappa,
\end{equation}
hence \eqref{claim2}.
\qed

\subsubsection{Linear stability}

\noindent
Let:
\begin{equation}
\mathsf{A}^{(\kappa)}\left(x,y\right) = p^{(\kappa)}\left(x\right)J\left(x,y\right), \qquad p^{(\kappa)}\left(x\right)=\beta\left[1-(m^{(\kappa)}\left(x\right))^2\right]
\end{equation}
and $\mathsf{A}^{(\kappa)}$ the operator with kernel $\mathsf{A}^{(\kappa)}\left(x,y\right)$. In \cite{DMOP} it is proved that for $\kappa$ small enough $\mathsf{A}^{(\kappa)}$ has a positive eigenvalue $\lambda^{(\kappa)}$ with eigenvector $u^{(\kappa)}$, bounded, in $L^1$ and positive: $u^{(\kappa)}\left(x\right)>0$ for all $x$. Moreover the rest of the spectrum of $\mathsf{A}^{(\kappa)}$ is in the interval $\left(-a,a\right)$ with $a<\lambda^{(\kappa)}$. Here we prove that:
\begin{prop}
For all $\kappa>0$:
\begin{eqnarray}
\lambda^{(\kappa)} &\le& 1-\kappa\tanh\left(\beta\kappa\right)\left(1-m_{\beta,\kappa}^2\right) \label{est_1} \\
\lambda^{(\kappa)} &\le& 1-{\kappa\over 2}\bar{m}\left(\delta\right)\left(1-m_{\beta,\kappa}^2\right)
\end{eqnarray}
$\delta>0$ small enough but independent of $\kappa$.
\end{prop}
\noindent
\textbf{Proof.} Since $\mathsf{A}^{(\kappa)}u^{(\kappa)} = \lambda^{(\kappa)}u^{(\kappa)}$:
\begin{equation}\label{evalue_n}
\big(\lambda^{(\kappa)}\big)^n u^{(\kappa)}\left(x_0\right)=\int\prod_{j=1}^n\,p^{(\kappa)}\left(x_{j-1}\right)J\left(x_{j-1},x_j\right)u^{(\kappa)}\left(x_n\right)\de x_j.
\end{equation}
We multiply and divide each term by $\bar{p}=\beta\left(1-\bar{m}^2\right)$. We have:
\begin{equation}
{p^{(\kappa)}\over\bar{p}}=1-{\left(m^{(\kappa)}-\bar{m}\right)\left(m^{(\kappa)}+\bar{m}\right)\over 1-\bar{m}^2}.
\end{equation}
By \eqref{claim2}:
\begin{equation}\label{bd_1}
\left|m^{(\kappa)}\left(x\right)-\bar{m}\left(x\right)\right|\ge\beta\kappa\left(1-m_{\beta,\kappa}^2\right)
\end{equation}
while
\begin{equation}
\left|m^{(\kappa)}\left(x\right)+\bar{m}\left(x\right)\right|\ge m_{\beta,\kappa}\left(0\right)=\tanh\left(\beta\kappa\right).
\end{equation}
Thus:
\begin{equation}\label{bd_2}
{p^{(\kappa)}\over\bar{p}}\le1-\beta\kappa\tanh\left(\beta\kappa\right)\left(1-m_{\beta,\kappa}^2\right)\eqqcolon\eta_{\kappa}
\end{equation}
then by \eqref{evalue_n}:
\begin{equation}\label{est_lam}
\big(\lambda^{(\kappa)}\big)^n u^{(\kappa)}\left(x\right)\le\eta_{\kappa}^n\mathsf{A}_{\bar{m}}^n u^{(\kappa)}\left(x\right)
\end{equation}
with $\mathsf{A}_{\bar{m}}\left(x,y\right)=\bar{p}\left(x\right)J\left(x,y\right)$. In \cite{DMOP} it is proved that:
\begin{equation}
\lim_{n\to\infty}\mathsf{A}_{\bar{m}}^n u^{(\kappa)}\left(x\right)=\bar{m}'\left(x\right){\left\langle\bar{m}',u^{(\kappa)}\right\rangle\over\left\langle\bar{m}',\bar{m}'\right\rangle},\qquad\left\langle f,g\right\rangle=\int{1\over 1-\bar{m}^2\left(x\right)}f\left(x\right)g\left(x\right)\de x
\end{equation}
hence \eqref{est_1} after taking the log in \eqref{est_lam}, dividing by $n$ and taking the limit $n\to\infty$.
\newline
The bound \eqref{est_1} is quadratic in $\kappa$, a bound linear in $\kappa$ can be obtained as follows. Instead of \eqref{bd_1} we bound:
\begin{equation}
\left|m^{(\kappa)}\left(x\right)+\bar{m}\left(x\right)\right|\ge\bar{m}\left(\delta\right),\qquad\left|x\right|\ge\delta
\end{equation}
and \eqref{bd_2} is then replaced by:
\begin{equation}
{p^{(\kappa)}\left(x\right)\over\bar{p}\left(x\right)}\le 1-a\mathbf{1}_{\left|x\right|\ge\delta},\qquad a\coloneqq \kappa\bar{m}\left(\delta\right)\left(1-m_{\beta,\kappa}^2\right)
\end{equation}
so that \eqref{evalue_n} becomes:
\begin{eqnarray}
\big(\lambda^{(\kappa)}\big)^n u^{(\kappa)}\left(x_0\right)&\le&\left(1-a\right)^n\int\prod_{j=1}^n\left(1-a\right)^{-\mathbf{1}_{|x_{j-1}|<\delta}}\bar{p}\left(x_{j-1}\right)J\left(x_{j-1},x_j\right)u^{(\kappa)}\left(x_n\right)\de x_j\nn \\
&=&\left(1-a\right)^n\int\prod_{j=1}^n\left(1-a\right)^{-\mathbf{1}_{|x_{j-1}|<\delta}}\mathsf{A}_{\bar{m}}\left(x_{j-1},x_j\right)u^{(\kappa)}\left(x_n\right)\de x_j.
\end{eqnarray}
We fix $s>0$ and $b>0$ so that:
\begin{equation}
\max_{\left|x\right|\ge s}\bar{p}\left(x\right)\ge\mathrm{e}^{-b}.
\end{equation}
We take $\left|x\right|<s$ and call:
\begin{equation}
\omega_{\kappa}\coloneqq\max_{\left|x\right|\ge s\text{, }y\in\mathbb{R}}{u^{(\kappa)}\left(y\right)\over u^{(\kappa)}\left(x\right)}.
\end{equation}
Given $x_0,\ldots,x_{n-1}$, call $k$ the largest integer such that $\left|x_{k}\right|\le s$ and $\left|x_{i}\right|>s$ for $i=k+1,\ldots,n-1$. Then:
\begin{equation}
\big(\lambda^{(\kappa)}\big)^n\le \omega_{\kappa}\left(1-a\right)^n\sum_{k=0}^{n-1}\int\prod_{j=1}^{k+1}\mathrm{e}^{-b\left(n-k\right)}\left(1-a\right)^{-\mathbf{1}_{|x_{j-1}|<\delta}}J\left(x_{j-1},x_j\right)\de x_j.
\end{equation}
Call:
\begin{equation}
\mathsf{Q}_{\bar{m}}\left(x,y\right)\coloneqq{\bar{m}'\left(y\right)\over\bar{m}'\left(x\right)}\,\mathsf{A}_{\bar{m}}\left(x,y\right)
\end{equation}
and $\mathbb{E}_x$ the expectation with respect to the Markov chain with transition probability $\mathsf{Q}_{\bar{m}}\left(x,y\right)$ which starts from $x$. Then:
\begin{equation}
\big(\lambda^{(\kappa)}\big)^n\le \omega\left(1-a\right)^n\sum_{k=0}^{n-1}\mathrm{e}^{-b\left(n-1-k\right)}\mathbb{E}_x\left[\prod_{j=1}^{k+1}\left(1-a\right)^{-\mathbf{1}_{|x_{j-1}|<\delta}}\right]
\end{equation}
where:
\begin{equation}
\omega\coloneqq\max_{\left|y\right|<s+1\text{, }\left|x\right|<s}{\bar{m}'\left(y\right)\over\bar{m}'\left(x\right)}.
\end{equation}
Let:
\begin{equation}
\pi_{\delta}\coloneqq\max_{x}\int_{-\delta}^{\delta}\mathsf{Q}_{\bar{m}}\left(x,y\right)\mathrm{d}y
\end{equation}
then:
\begin{equation}
\big(\lambda^{(\kappa)}\big)^n\le c\left(1-a\right)^n\sum_{k=0}^{n-1}\mathrm{e}^{-b\left(n-1-k\right)}\left(1-\pi_{\delta}+\pi_{\delta}\left(1-a\right)^{-1}\right)^{k+2}
\end{equation}
which gives:
\begin{equation}
\lambda^{(\kappa)}\le\log\left(1-a\right)+\log\left(1+\pi_{\delta}{a\over 1-a}\right)\le\log\left(1-{a\over 2}\right)
\end{equation}
if $\delta$ is small enough.
\qed
\newline
\newline
Theorem \ref{t1} is proved.
\qed

\section{Bounds on the starting profile}

\subsection{Introduction to the problem}

\noindent
The strategy we have in mind is an extension of that used in \cite{RB} to prove the existence of a solution for an unbiased stationary problem in a finite interval. Nevertheless, in that case boundary values share the same sign and the region $\left[0,m^*\left(\beta\right)\right]$ is completely avoided; this is not the case here so that a special caution is then needed when dealing with non-local equations of type
\begin{equation}\label{eqex}
\vf_{m,h}\left(x\right)-p_{m,h}\left(x\right)\left(\JB\star\vf_{m,h}\right)\left(x\right)=\tanh\Big\{\beta\Big[\left(\JB\star m\right)\left(x\right)+h\left(x\right)\Big]\Big\}-m\left(x\right)
\end{equation}
for any $x\in\IIp$, where $\vf_{m,h}$ is the unknown function, $m$ and $h$ are bounded in $\IIp$ and:
\begin{equation}
p_{m,h}\left(x\right)\coloneqq {\beta\over\cosh^2\Big\{\beta\Big[\left(\JB\star m\right)\left(x\right)+h\left(x\right)\Big]\Big\}}.
\end{equation}

\noindent
As we shall see, equation \eqref{eqex} plays a crucial role in our strategy. The problem arises here when inverting \eqref{eqex}. Formally, we would like to write:
\begin{equation}\label{non}
\vf_{m,h}\left(x\right)=\left(\mathsf{I}-\mathsf{A}_{m,h}\right)^{-1}\left[
\tanh\Big\{\beta\Big[\left(\JB\star m\right)\left(x\right)+h\left(x\right)\Big]\Big\}-m\left(x\right)\right]
\end{equation}
where $\mathsf{A}_{m,h}$ is a linear operator acting on functions in $L^{\infty}\left(\IIp\right)$ with kernel:
\begin{equation}
\mathsf{A}_{m,h}\left(x,y\right)\stackrel{\mathcal{D}'}{\coloneqq}p_{m,h}\left(x\right)\JB\left(x,y\right)
\end{equation}
for any $x\in\IIp$ and $y\in[-\vei-1,\vei+1]$. In fact $\mathsf{A}_{m,h}$ may be not invertible in this domain, so that expression \eqref{non} may not make sense. 
\newline
In \cite{EP} and \cite{DMOP} a general Perron-Frobenius theorem for $\mathsf{A}_{m,h}$ is proved, extension to our case is straightforward and then omitted. We rather provide here estimates on $\mathsf{A}_{m_0,h_0}$, $m_0$ as in \eqref{m0} and $h_0=\mathsf{T}^{(\kappa)}m_0$, which guarantee the existence of $\left(\mathsf{I}-\mathsf{A}_{m_0,h_0}\right)^{-1}$. This result directly enter the proof of the main theorem and will be recalled when needed. For computational convenience, we do not restrict in this section to the positive interval $(0,\vei]$. We indicate $\left\|\,\cdot\,\right\|_{\infty}\coloneqq\left\|\,\cdot\,\right\|_{L^{\infty}\left(\IIp\right)}$.

\subsection{Main estimates}

\noindent
Define
\begin{equation}
\eta_0\coloneqq\sup_{\left|x\right|<\veim,\left|y-x\right|\le 1}{p_{m_0,h_0}\left(x\right)\over\bar{p}\left(y\right)}, \qquad \pi_0\coloneqq \sup_{{\veim\over 2}\le\left|x\right|\le\vei}p_{m_0,h_0}\left(x\right).
\end{equation}
Notice that, by definition, $\pi_0<1$.
\begin{prop}\label{p41}
$\eta_0< 1$ provided $\ve$ is small enough.
\end{prop}
\noindent
\textbf{Proof.}
Take without loss of generality $x\in(0,\veim]$ and consider the sub-case $x\in(0,\veim-1]$. Expand in Taylor series:
\begin{eqnarray}\label{expt}
\tanh\Big\{\beta\Big[\left(\JB\star m_0\right)\left(x\right)+
h_0\left(x\right)\Big]\Big\}&=&\tanh\Big\{\beta\Big[\big(J\star m^{(\kappa)}\big)\left(x\right)+\kappa-j\ve\int_0^x \chi_{\beta}^{-1}\big(m^{(\kappa)}\left(y\right)\big)\de y\Big]\Big\}\nn\\
&=& m^{(\kappa)}\left(x\right)+p^{(\kappa)}\left(x\right)\left(-j\ve\int_0^x \chi_{\beta}^{-1}\big(m^{(\kappa)}\left(y\right)\big)\de y\right)\nn\\
&-&m^{(\kappa)}\left(x\right)p^{(\kappa)}\left(-j\ve\int_0^x \chi_{\beta}^{-1}\big(m^{(\kappa)}\left(y\right)\big)\de y\right)^2+\ldots
\end{eqnarray}
and observe that 
\begin{equation}
\sup_{\left|x\right|<\veim-1}\left|\,j\ve\int_0^x \chi_{\beta}^{-1}\big(m^{(\kappa)}\left(y\right)\big)\de y\right|\le \mathrm{const}\cdot \sqve.
\end{equation}
Hence
\begin{equation}
\tanh^2\Big\{\beta\Big[\left(\JB\star m_0\right)\left(x\right)+
h_0\left(x\right)\Big]\Big\}=\big[m^{(\kappa)}\left(x\right)\big]^2+O\left(\sqve\right),
\end{equation}
which implies
\begin{eqnarray}
\left|p_{m_0,h_0}\left(x\right)-p^{(\kappa)}\left(x\right)\right|=O\left(\sqve\right).
\end{eqnarray}
Therefore:
\begin{equation}\label{ineqp}
{p_{m_0,h_0}\left(x\right)\over\bar{p}\left(y\right)}\le{p^{(\kappa)}\left(x\right)\over\bar{p}\left(y\right)}+{\left|p_{m_0,h_0}\left(x\right)-p^{(\kappa)}\left(x\right)\right|\over\bar{p}\left(y\right)}\le\eta_{\kappa}+\mathrm{const}\cdot \sqve
\end{equation}
which is less than $1$ if $\ve$ is small enough. Conversely, if $x\in(\veim-1,\veim]$:
\begin{eqnarray}\label{stimm}
&&\tanh\Big\{\beta\Big[\left(\JB\star m_0\right)\left(x\right)+
h_0\left(x\right)\Big]\Big\}\nn\\
&=&\tanh\Big\{\beta\Big[\int_{x-1}^{\veim}J\left(x,y\right)m^{(\kappa)}\left(y\right)\de y +\int_{\veim}^{x+1}J\left(x,y\right)\hat{m}^{\mathrm{macro}}_{\ve}\left(y\right)\de y+h_0\left(x\right)\Big]\Big\}\nn\\
&=&\tanh\Big\{\beta\Big[\big(J\star m^{(\kappa)}\big)\left(x\right)+\kappa-j\ve\int_0^x \chi_{\beta}^{-1}\big(m^{(\kappa)}\left(y\right)\big)\de y\nn\\
&+&\int_{\veim}^{x+1}J\left(x,y\right)\left(\hat{m}^{\mathrm{macro}}_{\ve}\left(y\right)-m^{(\kappa)}\left(y\right)\right)\de y\Big]\Big\}.
\end{eqnarray}
In the interval $[\veim,x+1]$ $m^{(\kappa)}$ is increasing, while $\mmac$ may be decreasing or increasing depending on $\mu$. Consider the case $\mu\in\left(m^*\left(\beta\right),m_{\beta}\right)$, so that $\mmac$ is strictly decreasing, the complementary case being similar. The last integral in \eqref{stimm} can be bounded as follows:
\begin{eqnarray}\label{412}
&&\left|\int_{\veim}^{x+1}J\left(x,y\right)\left(\hat{m}^{\mathrm{macro}}_{\ve}\left(y\right)-m^{(\kappa)}\left(y\right)\right)\de y\right|\le\left|\mmac\left(x+1\right)-m^{(\kappa)}\left(x+1\right)\right|\nn\\
&\le&\left|\mmac\left(x+1\right)-\mmac(\veim)\right|+
\left|m^{(\kappa)}\left(x+1\right)-m^{(\kappa)}(\veim)\right|
\end{eqnarray}
where we used the fact that $\mmac(\veim)=m^{(\kappa)}(\veim)$. By \eqref{mmacro} and \eqref{m0} and using Lagrange's theorem:
\begin{equation}
\left|\mmac\left(x+1\right)-\mmac(\veim)\right|\le \sup_{\veim\le y\le x+1} \left|
{\de \mmac\over\de x}\left(y\right)\right|=O\left(\sqve\right).
\end{equation}
Furthermore, there are constants $c_{\kappa}$ and $\theta_{\kappa}>0$ such that
\begin{equation}
\left|m^{(\kappa)}\left(x+1\right)-m^{(\kappa)}(\veim)\right|\le\left|m^{(\kappa)}\left(x+1\right)-m_{\beta,\kappa}\right|+\left|m^{(\kappa)}(\veim)-m_{\beta,\kappa}\right|\le 2c_{\kappa} \mathrm{e}^{\theta_{\kappa}\veim}
\end{equation}
if $\ve$ is small enough, since $m^{(\kappa)}$ approaches $m_{\beta,\kappa}$ exponentially fast for $x\gg 1$.
Using the last estimates, we conclude that:
\begin{eqnarray}\label{415}
&&\tanh\Big\{\beta\Big[\left(\JB\star m_0\right)\left(x\right)+
h_0\left(x\right)\Big]\Big\}\nn\\
&\le&\tanh\Big\{\beta\Big[\big(J\star m^{(\kappa)}\big)\left(x\right)+\kappa-j\ve\int_0^x \chi_{\beta}^{-1}\big(m^{(\kappa)}\left(y\right)\big)\de y + \mathrm{const}\cdot\sqve\Big]\Big\}
\end{eqnarray}
for any $x\in(\veim-1,\veim]$. Performing the same expansion as in \eqref{expt}, we end up with an inequality similar to \eqref{ineqp}. In the end
\begin{equation}
\eta_0=\eta_{\kappa}+ O\left(\sqve\right)
\end{equation}
Proposition \ref{p41} is proved.
\qed
\begin{prop}
The following estimates hold for any $n\in\mathbb{N}$:
\begin{eqnarray}
\mathrm{(a)}\;\;\;&&\int_{\left|x_j\right|\le{\veim\over 2}}\prod_{j=1}^n\mathsf{A}_{m_0,h_0}\left(x_{j-1},x_j\right)\de x_j \le \eta_0^n, \qquad x_0\in\IIm\\
\mathrm{(b)}\;\;\;&&\int_{{\veim\over 2}<\left|x_j\right|\le\vei}\prod_{j=1}^n\mathsf{A}_{m_0,h_0}\left(x_{j-1},x_j\right)\de x_j\le\pi_0^n,\qquad x_0\in\Big[-\vei,-{\veim\over 2}\Big]\cup\Big[{\veim\over 2},\vei\Big].\nn\\
\end{eqnarray}
\end{prop}
\noindent
\textbf{Proof.} (b) is obtained combining the definition of $m_0$ with the fact that $\pi_0<1$. We prove (a). Let $x_0\in\IIm$ and write:
\begin{eqnarray}
\prod_{i=1}^n\mathsf{A}_{m_0,h_0}\left(x_{i-1},x_i\right)&=&\prod_{i=1}^n{p_{m_0,h_0}\left(x_{i-1}\right)\over \bar{p}\left(x_{i-1}\right)}\prod_{i=1}^n\mathsf{A}_{\bar{m}}\left(x_{i-1},x_i\right)\nn\\
&=&{\bar{m}'\left(x_0\right)\over\bar{m}'\left(x_n\right)}\prod_{i=1}^n{p_{m_0,h_0}\left(x_{i-1}\right)\over \bar{p}\left(x_{i-1}\right)}\prod_{i=1}^n\mathsf{Q}_{\bar{m}}\left(x_{i-1},x_i\right)\nn\\
&=&{\bar{m}'\left(x_n\right)\over\bar{m}'\left(x_0\right)}\prod_{i=1}^n{p_{m_0,h_0}\left(x_{i-1}\right)\over \bar{p}\left(x_{i}\right)}\prod_{i=1}^n\mathsf{P}_{\bar{m}}\left(x_i,x_{i-1}\right)\le\eta_0^n\,{\bar{m}'\left(x_n\right)\over\bar{m}'\left(x_0\right)}\prod_{i=1}^n\mathsf{P}_{\bar{m}}\left(x_i,x_{i-1}\right)\nn\\
\end{eqnarray}
where $\mathsf{P}_{\bar{m}}\left(y,x\right)$ is the inverse probability kernel which satisfies the detailed balance condition:
\begin{equation}
\mu\left(x\right)\mathsf{Q}_{\bar{m}}\left(x,y\right)=\mu\left(y\right)\mathsf{P}_{\bar{m}}\left(y,x\right)
\end{equation}
$\mu\left(\de x\right)$ being the invariant measure of the chain $\mathsf{Q}_{\bar{m}}$, $\mu\left(\mathrm{d}x\right)=\mu\left(x\right)\mathrm{d}x=\bar{C}\bar{p}^{-1}\left(x\right)\left(\bar{m}'\left(x\right)\right)^2\mathrm{d}x$, $\bar{C}$ a normalization constant (see \cite{EP}). Therefore:
\begin{equation}
\mathsf{Q}_{\bar{m}}\left(x,y\right)={\left(\bar{m}'\left(y\right)\right)^2\over\left(\bar{m}'\left(x\right)\right)^2}
{\bar{p}\left(x\right)\over\bar{p}\left(y\right)}\mathsf{P}_{\bar{m}}\left(y,x\right).
\end{equation}
In \cite{EP} it is also proved that there are constants $c_{\ve}>0$ and $\theta_{\ve}>0$ such that
\begin{equation}
\sup_{\left|x\right|\le{\veim\over 2},\left|y\right|\le{\veim\over 2}}{\bar{m}'\left(x\right)\over\bar{m}'\left(y\right)}\le c_{\ve}\mathrm{e}^{-\theta_{\ve}{\veim\over 2}},
\end{equation}
then
\begin{equation}
\int_{\left|x_j\right|\le{\veim\over 2}}\prod_{j=1}^n\mathsf{A}_{m_0,h_0}\left(x_{j-1},x_j\right)\de x_j \le c_{\ve}\mathrm{e}^{-\theta_{\ve}{\veim\over 2}}\eta_0^n\int_{\left|x_j\right|\le{\veim\over 2}}\prod_{j=1}^n\mathsf{P}_{\bar{m}}\left(x_{j-1},x_j\right)\de x_j \le\eta_0^n
\end{equation}
provided $\ve$ is small enough.
\qed
\newline
\newline
Let $\gamma_0\coloneqq\max\big\{\eta_0,\pi_0\big\}$; we have the following
\begin{prop}\label{propA}
There is a constant $b_0$, $0<b_0<1$, such that for any $n\in\mathbb{N}$:
\begin{equation}\label{azero}
\left\|\mathsf{A}_{m_0,h_0}^n\left(x_0,\,\cdot\,\right)\right\|_{\infty}\le {\left(\sqrt{\gamma_0}\right)^n\over 1-b_0}.
\end{equation}
\end{prop}
\noindent
\textbf{Proof.} Call:
\begin{equation}\label{lzero}
\mathcal{I}_n\left(x_0\right)\coloneqq\int\prod_{j=1}^n\mathsf{A}_{m_0,h_0}\left(x_{j-1},x_j\right)\de x_j, \qquad x_0\in\IIp
\end{equation}
and define sets:
\begin{eqnarray}
&&\Lambda_1\coloneqq\IIm\\
&&\Lambda_2\coloneqq\Big[-\vei,-{\veim\over 2}\Big]\cup\Big[{\veim\over 2},\vei\Big]\\
&&\Lambda_1^c\coloneqq\IIp\setminus\Lambda_1\\
&&\Lambda_2^c\coloneqq\IIp\setminus\Lambda_2\\
&&\Lambda_{1_{\mathrm{b}}}\coloneqq[-\veim-1,-\veim+1]\cup[\veim-1,\veim+1]\\
&&\Lambda_{2_{\mathrm{b}}}\coloneqq\Big[-{\veim\over 2}-1,-{\veim\over 2}+1\Big]\cup\Big[{\veim\over 2}-1,{\veim\over 2}+1\Big].
\end{eqnarray}
\noindent
Take without loss of generality $x_0\in\Lambda_1$ and $n$ even. Define the sequence of stopping times $\tau_1,\ldots,\tau_k$, $\tau_i\ge 0$ for any $i=1,\ldots,k$ in the following way: $\tau_1$ is the number of steps such that $x_1,\ldots,x_{\tau_1-1}\in\Lambda_1$ and $x_{\tau_1}\in\Lambda_{1_{\mathrm{b}}}$, $\tau_2$ is the number of steps such that $x_{\tau_1+1},\ldots,x_{\tau_1+\tau_2-1}\in\Lambda_2$ and $x_{\tau_1+\tau_2}\in\Lambda_{2_{\mathrm{b}}}$ and so on. Consider then the entropic decomposition which bounds \eqref{lzero}:
\begin{eqnarray}
&&\mathcal{I}_n\left(x_0\right)\le\sum_{k}\sum_{\tau_1,\ldots,\tau_k}\mathbf{1}_{\{\tau_1+\ldots+\tau_k\le n\}}
\int_{\Lambda_{1}^{\tau_{1}-1}}\prod_{j_1=1}^{\tau_1-1}\mathsf{A}_{m_0,h_0}\left(x_{j_1-1},x_{j_1}\right)\de x_{j_1}\int_{\Lambda_{1_{\mathrm{b}}}}\mathsf{A}_{m_0,h_0}\left(x_{\tau_1-1},x_{\tau_1}\right)\de x_{\tau_1}\nn\\
&&\int_{\Lambda_{1}^{\tau_{2}-1}}\prod_{j_2=\tau_1+1}^{\tau_1+\tau_2-1}\mathsf{A}_{m_0,h_0}\left(x_{j_2}-1,x_{j_2}\right)\de x_{j_2}\int_{\Lambda_{2_{\mathrm{b}}}}\mathsf{A}_{m_0,h_0}\left(x_{\tau_1+\tau_2-1},x_{\tau_1+\tau_2}\right)\de x_{\tau_1+\tau_2}\nn\\
\nn\\
&&\ldots\nn\\
\nn\\
&&\int_{\Lambda_{1}^{\tau_{k-1}-1}}\prod_{j_{k-1}=\tau_1+\ldots+\tau_{k-2}+1}^{\tau_1+\ldots+\tau_{k-1}-1}\mathsf{A}_{m_0,h_0}\left(x_{j_{k-1}-1},x_{j_{k-1}}\right)\de x_{j_{k-1}}\int_{\Lambda_{1_{\mathrm{b}}}}\mathsf{A}_{m_0,h_0}\left(x_{\tau_{k-1}-1},x_{\tau_{k-1}}\right)\de x_{\tau_{k-1}}\nn\\
&&\int_{\Lambda_{2}^{\tau_{k}-1}}\prod_{j_{k}=\tau_1+\ldots+\tau_{k-1}+1}^{\tau_1+\ldots+\tau_{k}-1}\mathsf{A}_{m_0,h_0}\left(x_{j_{k}-1},x_{j_{k}}\right)\de x_{j_{k}}\int_{\Lambda_{2_{\mathrm{b}}}}\mathsf{A}_{m_0,h_0}\left(x_{\tau_{k}-1},x_{\tau_{k}}\right)\de x_{\tau_{k}}\nn\\
&&\int_{\Lambda_1^{n-\tau_1-\ldots-\tau_n}}\prod_{j_{k+1}=\tau_1+\ldots+\tau_k+1}^{n}\mathsf{A}_{m_0,h_0}\left(x_{j_{k+1}-1},x_{j_{k+1}}\right)\de x_{j_{k+1}}\nn\\
&\le&\sum_{k}\sum_{\tau_1,\ldots,\tau_k}\mathbf{1}_{\{\tau_1+\ldots+\tau_k\le n\}}\,\gamma_0^n.
\end{eqnarray}
Call $\tau_{\text{min}}$ the integer part of $\veim/2$ and observe that it takes at least $\tau_{\text{min}}$ steps to make a jump from $\Lambda_1$ to $\Lambda_2$ or viceversa, thus:
\begin{eqnarray}
\mathcal{I}_n\left(x_0\right)&\le&\sum_{k}\sum_{\tau_1,\ldots,\tau_k}\mathbf{1}_{\{\tau_1+\ldots+\tau_k\le n\}}\mathbf{1}_{\{\tau_j\ge\tau_{\text{min}},\,j=1,\ldots,k\}}\,\gamma_0^n\nn\\
&=&\left(\sqrt{\gamma_0}\right)^n\sum_{k}\sum_{\tau_1,\ldots,\tau_k}\mathbf{1}_{\{\tau_1+\ldots+\tau_k\le n\}}\mathbf{1}_{\{\tau_j\ge\tau_{\text{min}},\,j=1,\ldots,k\}}\,\left(\sqrt{\gamma_0}\right)^{\tau_1}\ldots\left(\sqrt{\gamma_0}\right)^{\tau_k}\left(\sqrt{\gamma_0}\right)^{n-\tau_1-\ldots-\tau_k}\nn\\
&\le&\left(\sqrt{\gamma_0}\right)^n\sum_{k}\left(\sum_{\tau_1=\tau_{\text{min}}}^{n}\left(\sqrt{\gamma_0}\right)^{\tau_1}\right)\ldots\left(\sum_{\tau_k=\tau_{\text{min}}}^{n}\left(\sqrt{\gamma_0}\right)^{\tau_k}\right).
\end{eqnarray}
Considering that:
\begin{equation}\label{sereta}
\sum_{\tau=\tau_{\text{min}}}^n\left(\sqrt{\gamma_0}\right)^{\tau}={\left(\sqrt{\gamma_0}\right)^{\tau_{\text{min}}}-\left(\sqrt{\gamma_0}\right)^{n+1}\over 1-\sqrt{\gamma_0}}\le b_0
\end{equation}
for any integer $n$ large enough, with $b_0<1$ if $\tau_{\text{min}}$ is large enough (that is, if $\ve$ is suitably small), we have
\begin{equation}\label{sereta2}
\mathcal{I}_n\left(x_0\right)\le\left(\sqrt{\gamma_0}\right)^n\sum_k b_0^k\le{\left(\sqrt{\gamma_0}\right)^n\over 1-b_0}
\end{equation}
where $b_0$ actually depends on $\gamma_0$ and $\ve$. This proves the existence of the inverse of $\left(\mathsf{I}-\mathsf{A}_{m_0,h_0}\right)$ and that
\begin{equation}
\left(\mathsf{I}-\mathsf{A}_{m_0,h_0}\right)^{-1} \equiv \sum_{k=0}^{\infty}\mathsf{A}_{m_0,h_0}^k.
\end{equation}
\qed

\section{First iteration}

\noindent
From now on, we restrict to the positive domain $(0,\vei]$. This section is devoted to the proof of fundamental Lemmas, which will be exploited here to prove Proposition \ref{p.3} and later to complete the proof of Theorem 2.2. 
\newline
We construct $m_1$ as an infinite sum:
\begin{equation}
m_1\left(x\right)=m_0\left(x\right)+\sum_{j=1}^{\infty}\vf_{k}\left(x\right),\qquad x\in(0,\vei]
\end{equation}
where functions $\vf_{k}$, $k\in\mathbb{N}$, are defined further on. For notational convenience, call for any $x\in(0,\vei]$ and $k\in\mathbb{N}$:
\begin{eqnarray}
&&\phi_{k}\left(x\right)\coloneqq\sum_{j=1}^k\vf_{j}\left(x\right)\\
&&p_{k}\left(x\right)\coloneqq{\beta\over\cosh^2\Big\{\beta\Big[\left(\JB\star\left(m_{0}+\phi_{k-1}\right)\right)\left(x\right)+h_0\left(x\right)\Big]\Big\}}\\
&&\mathsf{A}_{k}\left(x,y\right)\stackrel{\mathcal{D}'}{\coloneqq}p_{k}\left(x\right)\JB\left(x,y\right)
\end{eqnarray}
with $\vf_{0}\equiv 0$, and $p_{0}\equiv p_{m_0,h_0}$, $\mathsf{A}_{0}\equiv\mathsf{A}_{m_0,h_0}$ by definition. 

\subsection{First correction to $m_0$}

\begin{prop}\label{vf1}
For any $\kappa>0$ there exists $\vf_{1}\in L^{\infty}\left((0,\vei]\right)$ which solves:
\begin{equation}\label{vf_1}
\vf_{1}\left(x\right)-p_{0}\left(x\right)\left(J_{\mathrm{b}}\star\vf_{1}\right)\left(x\right)=\tanh\Big\{\beta\Big[\left(\JB\star m_{0}\right)\left(x\right)+h_0\left(x\right)\Big]\Big\}-m_0\left(x\right)
\end{equation}
provided $\ve$ is small enough. Moreover, there is a constant $c_{0}>0$ such that $\left\|\vf_{1}\right\|_{\infty}\le c_{0}\sqve$.
\end{prop}
\noindent
\textbf{Proof.} First of all, we provide a uniform estimate for:
\begin{equation}
\Delta_{0,1}\left(x\right)\coloneqq\tanh\Big\{\beta\Big[\left(\JB\star m_{0}\right)\left(x\right)+h_0\left(x\right)\Big]\Big\}-m_0\left(x\right), \qquad x\in(0,\vei].
\end{equation}
We split the estimate in several parts.
\newline
If $0<x\le \veim-1$:
\begin{eqnarray}
\left|\Delta_{0,1}\left(x\right)\right|&=&\left|\tanh\Big\{\beta\Big[\big(J\star\mk\big)\left(x\right)+h_0\left(x\right)\Big]\Big\}-\mk\left(x\right)\right|\nn\\
&=&\left|\tanh\Big\{\beta\Big[\big(J\star\mk\big)\left(x\right)+h_0\left(x\right)\Big]\Big\}-\tanh\Big\{\beta\Big[\big(J\star\mk\big)\left(x\right)+\kappa\Big]\Big\}\right|\nn\\
&\le&\beta j\ve\int_0^x\chib^{-1}\left(m\left(y\right)\right)\de y\le c_1\sqve.
\end{eqnarray}
If $\veim+1\le x\le\vei$:
\begin{eqnarray}
\left|\Delta_{0,1}\left(x\right)\right|&=&\left|\tanh\Big\{\beta\Big[\big(\JB\star\mmac\big)\left(x\right)+h_0\left(x\right)\Big]\Big\}-\mmac\left(x\right)\right|\nn\\
&=&\left|\tanh\Big\{\beta\Big[\big(\JB\star\mmac\big)\left(x\right)+h_0\left(x\right)\Big]\Big\}-\tanh\Big\{\beta\Big[\mmac\left(x\right)+h_0\left(x\right)\Big]\Big\}\right|\nn\\
&\le&\beta\left|\big(\JB\star\mmac\big)\left(x\right)-\mmac\left(x\right)\right|\le\beta\max_{\left|y-x\right|<1}\left|\mmac\left(y\right)-\mmac\left(x\right)\right|\nn\\
&\le&\beta\left\|{\de\mmac\over\de x}\right\|_{\infty}\le c_2\ve
\end{eqnarray}
where we used Lagrange's Theorem and noticed that $(\vei-\veim)^{-1}\sim\ve$.
\newline
Let now $\veim-1<x<\veim+1$ and consider the sub-case $\veim-1<x<\veim$. We have:
\begin{eqnarray}
\left|\Delta_{0,1}\left(x\right)\right|&\le&\Big|\tanh\Big\{\beta\Big[\int_{x-1}^{\veim}J\left(x,y\right)\mk\left(y\right)\de y+\int_{\veim}^{x+1}J\left(x,y\right)\mmac\left(y\right)\de y +h_0\left(x\right)\Big]\Big\}\nn\\
&-&\mk\left(x\right)\Big|\le\beta\int_{\veim}^{x+1}J\left(x,y\right)\left|\mmac\left(y\right)-\mk\left(y\right)\right|\de y+c_1\sqve\nn\\
\end{eqnarray}
Recalling the definition of $m_0$, $\mk\,(\veim)=\mmac\,(\veim)$, thus:
\begin{equation}
\left|\Delta_{0,1}\left(x\right)\right|\le\beta\left|\mmac\left(x+1\right)-\mmac\,(\veim)\right|+\beta\left|\mk\left(x+1\right)-\mk\,(\veim)\right|+c_1\sqve\le c_3\sqve
\end{equation}
by virtue of estimates \eqref{412}$-$\eqref{415}.
\newline
The last case we have to take into account is when $\veim<x<\veim+1$. We have:
\begin{eqnarray}
\left|\Delta_{0,1}\left(x\right)\right|&\le&\Big|\tanh\Big\{\beta\Big[\int_{x-1}^{\veim}J\left(x,y\right)\mk\left(y\right)\de y+\int_{\veim}^{x+1}J\left(x,y\right)\mmac\left(y\right)\de y +h_0\left(x\right)\Big]\Big\}\nn\\
&-&\mmac\left(x\right)\Big|\le\beta\int_{x-1}^{\veim}\left[J\left(x,y\right)\mk\left(y\right)-\mmac\left(y\right)\right]\de y +c_2\ve\nn\\
&\le&\beta\left|\mk\left(x-1\right)-\mmac\left(x-1\right)\right|+c_2\ve
\end{eqnarray}
where we bound the first term in the last row as before. Collecting the previous estimates, we deduce that there exist $c'>0$ such that $\left\|\Delta_{0,1}\right\|_{\infty}\le c'_0\sqve$.
\newline
This is all we need in order to prove Proposition \ref{vf1}. In fact by Proposition \ref{propA}, $\vf_{1}$ does exist because $\left(\mathsf{I}-\mathsf{A}_{0}\right)$ can be inverted, and then $\left\|\vf_{1}\right\|_{\infty}\le c_{0}\sqve$, $c_0\coloneqq c'_0\left(1-b_0\right)^{-1}\left(1-\sqrt{\gamma_0}\right)^{-1}$.
\qed

\subsection{Fundamental results}

\begin{lem}\label{lemF}
There exists $\delta_0>0$ such that for any $\delta<\delta_0$ the following holds: for any $F\in L^{\infty}\left((0,\vei]\right)$, if $\left\|p-p_{0}\right\|_{\infty}<\delta$, the equation
\begin{equation}\label{eqnF}
\vf\left(x\right)-p\left(x\right)\left(\JB\star \vf\right)\left(x\right) = F\left(x\right), \qquad x\in(0,\vei]
\end{equation}
can be solved in the unknown function $\vf$. Moreover, there exists a constant $c_{\delta}>0$ such that 
\begin{equation}\label{pippo1}
\left\|\vf\right\|_{\infty}\le c_{\delta}\left\|F\right\|_{\infty}.
\end{equation}
\end{lem}
\noindent
\textbf{Proof.}
Call
\begin{equation}
\eta'\coloneqq \sup_{\left|x\right|\le\veim,\left|y-x\right|\le 1} {p\left(x\right)\over \bar{p}\left(y\right)}, \qquad \pi'\coloneqq\sup_{\veim\le\left|x\right|\le \vei} p\left(x\right).
\end{equation}
We easily have, for any $x\in(0,\veim]$ and $y$ such that $\left|y-x\right|\le 1$:
\begin{equation}\label{eta30}
\eta' \le \sup_{\left|x\right|\le\veim,\left|y-x\right|\le 1}\left({p_{0}\left(x\right)\over \bar{p}\left(y\right)}+{p\left(x\right)-p_{0}\left(x\right)\over \bar{p}\left(y\right)} \right)\le \eta_0+\chi_{\beta}^{-1}\left(m_{\beta}\right)\delta.
\end{equation}
Similarly, we get $\pi'\le \pi_0+\delta$. Call $\gamma'\coloneqq \max\left\{\eta',\pi'\right\}$ and repeat the strategy used to prove \eqref{lzero} taking $\ve$ small enough. Then, \eqref{pippo1} follows immediately with $c_{\delta}$ which bounds the $L^{\infty}$ norm of $\left(\mathsf{I}-p\,\JB\,\star\,\right)^{-1}$. Notice that by \eqref{eta30}, $c_{\delta}<c_{\delta_0}$ for any $\delta<\delta_0$.
\newline
\qed
\begin{lem}\label{facile}
For any $m, h$ and $\phi$ in $L^{\infty}\left((0,\vei]\right)$:
\begin{equation}
\left\|p_{m+\phi,h}-p_{m,\phi}\right\|_{\infty}\le 2\beta^2 \left\|\phi\right\|_{\infty}.
\end{equation}
\end{lem}
\noindent
\textbf{Proof.} Since for any $z_1$, $z_2\in\mathbb{R}$:
\begin{equation}\label{gen}
\left|\tanh^2\left(z_2\right)-\tanh^2\left(z_1\right)\right|\le 2 \tanh\left|z_2-z_1\right|,
\end{equation}
we get for any $x\in(0,\vei]$:
\begin{equation}
\left|p_{m+\phi,h}\left(x\right)-p_{m,\phi}\left(x\right)\right|\le 2\beta\tanh\Big\{\beta\Big[\big(J\star\left|\phi\right|\big)\left(x\right)\Big]\Big\}\le 2\beta^2\left\|\phi\right\|_{\infty}.
\end{equation}
\qed

\subsection{Convergence to $m_1$}

\begin{prop}\label{29set}
For any $\kappa>0$ there is $\ve^*>0$ such that for any $\ve<\ve^*$ the following holds:
\newline
\newline
$\mathrm{(i)}$ there exists $\vf_{k}\in L^{\infty}\left((0,\vei]\right)$ which solves
\begin{equation}\label{vfk}
\vf_{k}\left(x\right)-p_{k-1}\left(x\right)\left(J_{\mathrm{b}}\star\vf_{k}\right)\left(x\right)=\tanh\Big\{\beta\Big[\left(\JB\star\left(m_{0}+\phi_{k-1}\right)\right)\left(x\right)+h_0\left(x\right)\Big]\Big\}
-m_0\left(x\right)-\phi_{k-1}\left(x\right),
\end{equation}
$x\in(0,\vei]$, for any $k\ge 1$, where $\vf_{1}$ solves \eqref{vf1};
\newline
\newline
$\mathrm{(ii)}$
\begin{equation}
\lim_{k\to\infty}\left\|m_0+\phi_{k}\right\|_{\infty} = m_1
\end{equation}
where $m_1$ solves \eqref{m1};
\newline
\newline
$\mathrm{(iii)}$ there is a constant $c>0$ such that
\begin{equation}
\left\|m_1-m_0\right\|_{\infty}\le c\sqve.
\end{equation}
\end{prop}
\noindent
\textbf{Proof.} It works by iteration. We suppose that for any $k\le\bar{k}$, $\bar{k}$ fixed, there exists $\vf_{k}$ that solves \eqref{vfk} and that
\begin{equation}\label{r0}
\left\|\vf_{k}\right\|_{\infty}\le r_0\left\|\vf_{k-1}\right\|_{\infty}^2, \qquad 2\le k\le \bar{k}.
\end{equation}
where $r_0\coloneqq \beta c_{\delta_0}$. In this hypothesis, by iteration of \eqref{r0} and recalling that $\left\|\vf_{1}\right\|_{\infty}\le c_0\sqve$, we get
\begin{equation}\label{30}
\left\|\vf_{k}\right\|_{\infty}\le r_0^{2^{k-1}-1}\left(c_0\sqve\right)^{2^{k-1}}
\end{equation}
thus, for any $2\le k\le \bar{k}$
\begin{equation}\label{29.2}
\left\|\phi_{k-1}\right\|_{\infty}\le \sum_{j=1}^{k-1} r_0^{2^{j-1}-1}\left(c_0\sqve\right)^{2^{j-1}}\le {1\over r_0}\sum_{j=0}^{\infty} \left(c_0 r_0\sqve\right)^{2^j}\le{1\over r_0}\sum_{j=1}^{\infty}\left(c_0 r_0\sqve\right)^j
\le 2 c_0\sqve 
\end{equation}
provided $\ve<\left(2 c_0 r_0\right)^{-2}$. By Lemma \ref{facile}, we have:
\begin{equation}
\left\| p_{k-1}-p_{0}\right\|_{\infty}\le 2\beta^2 \left\|\phi_{k-1}\right\|_{\infty}\le 4\beta^2 c_0\sqve.
\end{equation}
Therefore, if we choose $\ve<\left(\delta/4\beta^2 c_0\right)^2$ we are in the hypothesis of Lemma \ref{lemF}, that guarantees the existence of $\vf_{\bar{k}+1}$. It remains to prove that \eqref{r0} holds for $\bar{k}+1$. Expand the hyperbolic tangent to get
\begin{eqnarray}\label{buh}
\vf_{\bar{k}+1}\left(x\right)-p_{\bar{k}}\left(x\right)\left(\JB\star\vf_{\bar{k}+1}\right)\left(x\right)&=&
\tanh\Big\{\beta\Big[\left(\JB\star\left(m_{0}+\phi_{\bar{k}-1}\right)\right)\left(x\right)+h_0\left(x\right)\Big]\Big\}\nn\\
&-&m_0\left(x\right)-\phi_{\bar{k}-1}\left(x\right)\vf_{\bar{k}}\left(x\right)+p_{\bar{k}-1}\left(x\right)\left(\JB\star\vf_{\bar{k}}\right)\left(x\right)\nn\\
&+&p'_{\bar{k}-1}\left(x\right)\left(\JB\star\vf_{\bar{k}}\right)^2\left(x\right)+\ldots
\end{eqnarray}
where, in general:
\begin{equation}
p_{k}'\left(x\right)\coloneqq p_{k}\left(x\right)\tanh\Big\{\beta\Big[\left(\JB\star\left(m_{0}+\phi_{k}\right)\right)\left(x\right)+h_0\left(x\right)\Big]\Big\}, \qquad k\in\mathbb{N}.
\end{equation}
By definition of $\vf_{\bar{k}}$, \eqref{buh} becomes:
\begin{equation}
\vf_{\bar{k}+1}\left(x\right)-p_{\bar{k}}\left(x\right)\left(\JB\star\vf_{\bar{k}+1}\right)\left(x\right)=p'_{\bar{k}-1}\left(x\right)\left(\JB\star\vf_{\bar{k}}\right)^2\left(x\right)+\ldots
\end{equation}
then:
\begin{equation}
\left\|\vf_{\bar{k}+1}\right\|_{\infty}\le c_{\delta} \sup_{0<x'\le\vei}\left|p'_{\bar{k}-1}\left(x'\right)\right|\sup_{0<x''\le\vei}\vf_{\bar{k}-1}^2\left(x''\right)\le \beta c_{\delta} \left\|\vf_{\bar{k}}\right\|^2_{\infty}.
\end{equation}
(i) is then proved. (ii) is a straightforward consequence of Newton's method, indeed:
\begin{equation}
m_0\left(x\right)+\phi_{k}\left(x\right)=\tanh\Big\{\beta\Big[\left(\JB\star\left(m_{0}+\phi_{k}\right)\right)\left(x\right)+h_0\left(x\right)\Big]\Big\}+O\left(\vf_{k-1}^2\left(x\right)\right)
\end{equation}
for any $x\in(0,\vei]$ where, according to \eqref{30}, $\vf_{k}$ goes uniformly to zero as $\ve^{2^{k-1}}$.
(iii) comes from estimate \eqref{29.2}. Proposition \ref{29set} is proved. Observe that $m_1\in C^1\left((0,\vei]\right)$ since $J$ is smooth by hypothesis, $h_0$ is differentiable and $m_0$ continuous. Proposition \ref{p.3} is then proved.
\newline
\qed
\begin{cor}\label{coro}
There is a constant $c'>0$ such that
\begin{equation}
\left\|h_1-h_0\right\|_{\infty}\le c'\left\|m_1-m_0\right\|_{\infty}.
\end{equation}
\end{cor}
\noindent
\textbf{Proof.} By \eqref{gen} 
\begin{equation}
\left\|\chi_{\beta}\left(m_1\right)-\chi_{\beta}\left(m_0\right)\right\|_{\infty}\le 2\beta^2\left\|m_1-m_0\right\|_{\infty}.
\end{equation}
Call $s_0\coloneqq \inf_{0<x\le\vei}\chi_{\beta}\left(m_0\left(x\right)\right)$. We have:
\begin{equation}
\left\|\chi_{\beta}^{-1}\left(m_1\right)\chi_{\beta}^{-1}\left(m_0\right)\right\|_{\infty}\le\Big(
s_0\left(s_0-\left\|\chi_{\beta}\left(m_1\right)-\chi_{\beta}\left(m_0\right)\right\|_{\infty}\right)
\Big)^{-1}\eqqcolon \zeta_0
\end{equation}
so that, since
\begin{equation}
\left|h_1\left(x\right)-h_0\left(x\right)\right|\le \left|j\right|\ve \int_0^x {\left|m_1^2\left(y\right)-m_0^2\left(y\right)\right|\over\beta\left(1-m_0^2\left(y\right)\right)\left(1-m_1^2\left(y\right)\right)}\de y\nn\\,
\end{equation}
we get
\begin{equation}
\left\|h_1-h_0\right\|_{\infty}\le 2\beta^2\left|j\right|\zeta_0 \left\|m_1-m_0\right\|_{\infty}, 
\end{equation}
thus $c'\coloneqq 2\beta^2\left|j\right|\zeta_0$.
\newline
\qed

\section{Convergence to the mesoscopic profile}

\subsection{Preliminaries, notation}

\noindent
This section is devoted to the proof of Proposition \ref{p.4}. We show that the strategy we suggested to prove existence for the first iterate can be opportunely replied to prove convergence for any iterate. In order to do this, we exploit scaling properties of the magnetization profiles by introducing a weighted norm:
\begin{equation}
\left\|m\right\|_{\alpha}\coloneqq\sup_{0<x\le\vei}\mathrm{e}^{-\alpha\ve x}\left|m\left(x\right)\right|,\qquad m\in L^{\infty}\left((0,\vei]\right).
\end{equation}
where $\alpha>0$ is a fixed parameter. Existence of iterates directly follows from the fact that the map $m_{n}\mapsto m_{n+1}$ (the same for $h_n$) is a contraction in the $\alpha$-norm for a feasible choice of $\alpha$ and $\ve$. Such property will be also the key to prove that the sequence $\left(m_n,h_n\right)_{n=0}^{\infty}$ uniformly converges for $n\to\infty$ to a solution of \eqref{systm}. Notice that $\left\|\,\cdot\,\right\|_{\alpha}\le\left\|\,\cdot\,\right\|_{\infty}\le \mathrm{e}^{\alpha} \left\|\,\cdot\,\right\|_{\alpha}$. We prove at first some fundamental results.

\subsection{Small perturbations to $m_0$}

\begin{lem}\label{l1}
There exists $\delta^*_0>0$ such that for any $\delta^*<\delta^*_0$, if $h\in C^1\left((0,\vei]\right)$ satisfies $\left\|h-h_0\right\|\le \delta^*$, there is $m\in C^1\left((0,\vei]\right)$ which solves 
\begin{equation}
m\left(x\right)=\tanh\Big\{\beta\Big[\left(\JB\star m\right)\left(x\right)+h\left(x\right)\Big]\Big\}.
\end{equation}
Moreover:
\begin{equation}
\left\|m-m_0\right\|_{\infty} \le \mathrm{const}\cdot \left\|h-h_0\right\|_{\infty}, \qquad
\left\|p-p_{0}\right\|_{\infty} \le \mathrm{const}\cdot \left\|h-h_0\right\|_{\infty},
\end{equation}
where $p\equiv p_{m,h}$.
\end{lem}
\noindent
\textbf{Proof.} We again make use of Newton's method starting from the pair $\left(m_0,h\right)$. Call $\hat{p}_{0}\equiv p_{m_0,h}$; using \eqref{gen}, we obtain 
\begin{equation}
\left\|\hat{p}_{0}-p_{0}\right\|_{\infty}\le 2\beta\tanh\Big\{\beta\left\|h-h_0\right\|_{\infty}\Big\}\le 2\beta^2\delta^*.
\end{equation}
By Lemma \ref{lemF}, if $\delta^*_0<\delta_0/2\beta^2$ there exists $\psi_{1}\in L^{\infty}\left((0,\vei]\right)$ solution of
\begin{equation}
\psi_{1}\left(x\right)-\hat{p}_{0}\left(x\right)\left(J_{\mathrm{b}}\star\psi_{1}\right)\left(x\right)=\tanh\Big\{\beta\Big[\left(\JB\star m_{0}\right)\left(x\right)+h\left(x\right)\Big]\Big\}-m_0\left(x\right)
\end{equation}
with $\left\|\psi_{1}\right\|_{\infty}\le r_0\left\|h-h_0\right\|_{\infty}$. Again, the purpose is to construct $m$ as 
\begin{equation}
m\left(x\right)=m_0\left(x\right)+\sum_{j=1}^{\infty}\psi_{k}\left(x\right), \qquad x\in(0,\vei]
\end{equation}
where the $\psi_{k}$'s are solution of
\begin{equation}\label{psik}
\psi_{k}\left(x\right)-\hat{p}_{k-1}\left(x\right)\left(\JB\star\vf_{k}\right)\left(x\right)=\tanh\Big\{\beta\Big[\Big(\JB\star\big(m_{0}+\Psi_{k-1}\big)\Big)\left(x\right)+h_0\left(x\right)\Big]\Big\}
-m_0\left(x\right)-\Psi_{k-1}\left(x\right)
\end{equation}
with $\Psi_{k}\coloneqq\sum_{j=1}^{k}\psi_{j}$ and $\hat{p}_{k}\equiv p_{m_0+\Psi_{k},h}$ for any $k\ge 1$. By induction, we can prove in a way similar at all to that used to prove Proposition \ref{29set} that the scheme works provided $\delta^*\le 2\beta^2\left(2r_0+1\right)$ and $\ve$ is small enough. Indeed, in this position we are in the hypothesis of Lemma \ref{lemF}, therefore the $\psi_{k}$'s are well defined for any $k\ge 1$ and $\left\|\psi_{k+1}\right\|_{\infty}=O\big(\left\|\psi_{k}\right\|_{\infty}^2\big)$. The uniform convergence to $m$ follows again by definition of Newton's method. 
\newline
\qed
\begin{lem}\label{lemmaalpha}
Let $h'$ and $h''$ such that $\left\|h'-h_0\right\|_{\infty}\le\delta^*$ and $\left\|h''-h_0\right\|_{\infty}\le\delta^*$, $\delta^*<\delta^*_0$.
Then, at fixed $\alpha>0$:
\begin{equation}\label{615}
\left\|m''-m'\right\|_{\alpha}\le r_0 \left\|h''-h'\right\|_{\alpha}
\end{equation}
where $m'$ and $m''$ respectively solve in $(0,\vei]$:
\begin{eqnarray}
m'\left(x\right)&=&\tanh\Big\{\beta\Big[\left(\JB\star m'\right)\left(x\right)+h'\left(x\right)\Big]\Big\}\\
m''\left(x\right)&=&\tanh\Big\{\beta\Big[\left(\JB\star m''\right)\left(x\right)+h''\left(x\right)\Big]\Big\}
\end{eqnarray}
provided $\ve$ is small enough.
\end{lem}
\noindent
\textbf{Proof.} Existence of $m'$ and $m''$ follows from Lemma \ref{l1}. Write
\begin{equation}
m''\left(x\right)=\tanh\Big\{\beta\Big[\left(\JB\star m'\right)\left(x\right)+h'\left(x\right)\Big]\Big\}
+p^*\left(x\right)\Big(\left(\JB\star \left(m''-m'\right)\right)\left(x\right)+\left(h''\left(x\right)-h'\left(x\right)\right)
\Big)
\end{equation}
where $p^*$ is an intermediate value between $p_{m',h'}$ and $p_{m'',h''}$. We have then
\begin{equation}
m''\left(x\right)-m'\left(x\right)-p^*\left(x\right)\left(\JB\star \left(m''-m'\right)\right)\left(x\right)=
p^*\left(x\right)\left(h''\left(x\right)-h'\left(x\right)\right);
\end{equation}
multiplying both members by $\mathrm{e}^{-\alpha\ve x}$ and taking the absolute values we get
\begin{equation}\label{1ott}
\mathrm{e}^{-\alpha\ve x}\left|m''\left(x\right)-m'\left(x\right)\right|-p^*\left(x\right)\mathrm{e}^{\alpha\ve}
\left(\JB\star \left|m''-m'\right|\right)\left(x\right)\le p^*\left(x\right)\mathrm{e}^{-\alpha\ve x}\left|h''\left(x\right)-h'\left(x\right)\right|.
\end{equation}
We know that if $\left\|p^*\mathrm{e}^{\alpha\ve}-p_{0}\right\|_{\infty}<\delta_0$, the corresponding equality in \eqref{1ott} can be solved in the unknown function $\mathrm{e}^{-\alpha\ve x}\left|m''\left(x\right)-m'\left(x\right)\right|$ and that:
\begin{equation}
\left\|m''-m'\right\|_{\alpha}\le c_{\delta_0}\left\|p^*\right\|_{\infty}\left\|h''-h'\right\|_{\alpha}\le
r_0 \left\|h''-h'\right\|_{\alpha}.
\end{equation}
Choose then $\delta^*_0$ small enough depending on the values of $\alpha$ and $\ve$.
\newline
\qed

\subsection{Proof of Proposition \ref{p.4}}
\noindent
\textbf{Proof.} The proof works by induction. Suppose that for any $n\le\bar{n}$, $\bar{n}\ge 2$ fixed, there exists $m_n\in C^1\left((0,\vei]\right)$ which solves
\begin{equation}
m_n\left(x\right)=\tanh\Big\{\beta\Big[\left(\JB\star m_n\right)\left(x\right)+
h_{n-1}\left(x\right)\Big]\Big\},
\end{equation}
$m_0$ as in \eqref{m0}. Moreover, suppose that there is $\rho\in\left(0,1\right)$ such that
\begin{equation}
\left\|h_n-h_{n-1}\right\|_{\alpha}\le \rho \left\|h_{n-1}-h_{n-2}\right\|_{\alpha}, \qquad 2\le n\le\bar{n}.
\end{equation}
In this hypothesis
\begin{equation}
\left\|h_n-h_{n-1}\right\|_{\alpha}\le \left\|h_1-h_0\right\|_{\alpha}\sum_{j=1}^{n-1}\rho^j\le {\rho\over 1-\rho} \left\|h_1-h_0\right\|_{\alpha}
\end{equation}
and
\begin{equation}\label{bd1ot}
\left\|h_n-h_0\right\|_{\alpha}\le{\rho\over 1-\rho} \left\|h_1-h_0\right\|_{\alpha}.
\end{equation}
\eqref{bd1ot} guarantees that we are in the hypothesis of Lemma \ref{29set}, so we conclude that there exists $m_{\bar{n}+1}\in C^1\left((0,\vei]\right)$ solution of
\begin{equation}
m_{\bar{n}+1}\left(x\right)=\tanh\Big\{\beta\Big[\left(\JB\star m_{\bar{n}+1}\right)\left(x\right)+
h_{\bar{n}}\left(x\right)\Big]\Big\}.
\end{equation}
Let $\zeta>0$ bound $\left\|\chi_{\beta}^{-1}\left(m_n\right)\chi_{\beta}^{-1}\left(m_{n-1}\right) \right\|_{\infty}$ for any $2\le n\le\bar{n}$. Performing the same estimate as in Corollary \ref{coro}, we get:
\begin{eqnarray}
\left|h_{\bar{n}+1}\left(x\right)-h_{\bar{n}}\left(x\right)\right|&\le& \left(2\beta^2\left|j\right|\zeta\right)\ve\int_{0}^x \mathrm{e}^{-\alpha\ve y}\left|m_{\bar{n}+1}\left(y\right)-m_{\bar{n}}\left(y\right)\right|
\mathrm{e}^{\alpha\ve y}\de y\nn\\
&\le&c''\ve\left\|m_{\bar{n}+1}-m_{\bar{n}}\right\|_{\alpha}\int_0^x\mathrm{e}^{\alpha\ve y}\de y\nn\\
&\le& {c''\mathrm{e}^{\alpha\ve x}\over\alpha}\left\|m_{\bar{n}+1}-m_{\bar{n}}\right\|_{\alpha}
\end{eqnarray}
where $c''\coloneqq 2\beta^2\left|j\right|\zeta$. Multiplying both members by $\mathrm{e}^{-\alpha\ve x}$ and taking the supremum with respect to $x\in(0,\vei]$ we get
\begin{equation}
\left\|h_{\bar{n}+1}-h_{\bar{n}}\right\|_{\alpha}\le {c''\over\alpha}\left\|m_{\bar{n}+1}-m_{\bar{n}}\right\|_{\alpha}\le {r_0 c''\over\alpha}\left\|h_{\bar{n}}-h_{\bar{n}-1}\right\|_{\alpha}
\end{equation}
where we used \eqref{615}. If $\alpha$ is larger than $r_0 c''$, we can take $\rho\coloneqq r_0 c''$. Thus, the application which maps $h_{n}$ to $h_{n+1}$ is a contraction in the $\alpha$-norm, the same for $m_n$. This implies the uniform convergence of $\left(m_n,h_n\right)_{n\in\mathbb{N}}$ to $\left(m,h\right)$ solution of problem \eqref{systm} for $\ve$ small enough. Moreover, $m\in C^1\left((0,\vei]\right)$ since $h$ is differentiable and $m$ continuous. 
\newline
\qed

\section{Invertibility of the scheme}

\noindent
We just proved that, at fixed $j$, a solution to \eqref{systm} can be obtained in an iterative way. This solution satisfies a certain boundary condition, namely $m\,(\vei)=\nu$, $\nu\in\mathscr{M}_{\beta}$, which may be different with respect to the starting one $m^{\mathrm{macro}}\left(1\right)=\mu$, although $\left|\mu-\nu\right|=O\left(\sqve\right)$. The purpose in this section is to prove that for any $\mu\in\mathscr{M}_{\beta}$ there is at least one $j$ such that $m_0$ is mapped to $m$, with $m\,(\vei)=\mu$. This is so because as a function of $x$ and $j$, $m\left(x;j\right)$ is uniformly continuous in the parameter $j$, as we shall prove below, and we can apply the Intermediate Value Thoerem. This is not enough to guarantee the local invertibility of $m$ with respect to $j$, which would require a bound on the possible derivatives of $m$ with respect to $j$. Nevertheless, we are not actually interested here in the uniqueness issue, since numerical simulations strongly suggest that \textit{bump-shaped} profile do exist besides antisymmetric solutions (see \cite{CDMP} for the case $\kappa=0$). Indeed, we prove here more than uniform continuity, as we show that $m\left(x;j\right)$ is Lipschitz in $j$.

\subsection{Lipschitz continuity for the starting profile}

\begin{prop}
There are constants $\gamma>0$, $\gamma'>0$ such that
\begin{equation}\label{Dj}
\left\|m_0\left(\,\cdot\,;j\right)-m_0\left(\,\cdot\,;j'\right)\right\|_{\infty}\le \gamma\left| j-j'\right|, \qquad \left\|h_0\left(\,\cdot\,;j\right)-h_0\left(\,\cdot\,;j'\right)\right\|_{\infty}\le \gamma'\left| j-j'\right|.
\end{equation}
\end{prop}
\noindent
\textbf{Proof.} $m_0$ and $h_0$ are differentiable with respect to $j$. Observing that in $(0,\vei]$ $m_0$ does not depend on $j$, we get by \eqref{jr}
\begin{equation}\label{0m}
{\partial\over\partial j} m_0\left(x;j\right)=
\begin{dcases}
0\qquad &x\in(0,\veim)\\
-{1\over 1-\chi_{\beta}\left(m_0\left(x;j\right)\right)}{\ve x-\sqve\over 1-\sqve} &x\in[\veim,\vei].
\end{dcases}
\end{equation}
Since ${\partial\over\partial j}  m_0\,(\veim;j)=0$, ${\partial\over\partial j}  m_0\left(x;j\right)$ is continuous for any $x\in(0,\vei]$. Concerning the magnetic field, we get
\begin{equation}\label{0h}
{\partial\over\partial j} h_0\left(x;j\right)=
\begin{dcases}
0\qquad &x\in(0,\veim)\\
-{1\over\chi_{\beta}\left(m_0\left(x;j\right)\right)}{\ve x-\sqve\over 1-\sqve}&x\in[\veim,\vei].
\end{dcases}
\end{equation}
Call $\chi^+\left(j\right)\coloneqq \sup_{0<x\le\vei}\chi_{\beta}\left(m_0\left(x;j\right)\right)$ and  $\chi^-\left(j\right)\coloneqq \inf_{0<x\le\vei}\chi_{\beta}\left(m_0\left(x;j\right)\right)$. Then, define $\gamma\coloneqq\max\left\{1/1-\chi^+\left(j\right),1/1-\chi^+\left(j'\right)\right\}$ and $\gamma'\coloneqq\max\left\{1/\chi^+\left(j\right),1/\chi^+\left(j'\right)\right\}$ and use Lagrange's Theorem to get \eqref{Dj}.
\newline
\qed

\begin{comment}
\begin{prop}
There are constants $\gamma>0$, $\gamma'>0$ such that
\begin{equation}
\sup_{0<x\le\vei}\left|m_1\left(x;j\right)-m_1\left(x;j'\right)\right|\le \gamma_0\left| j-j'\right|, \qquad \sup_{0<x\le\vei}\left|h_1\left(x;j\right)-h_1\left(x;j'\right)\right|\le \gamma'_0\left| j-j'\right|.
\end{equation}
\end{prop}
\noindent
\textbf{Proof.} We have
\begin{equation}\label{75}
\left|m_1\left(x;j\right)-m_1\left(x;j'\right)\right| \le \tilde{p}\left(x\right)\Big(\JB\star\left|m_1\left(x;j\right)-m_1\left(x;j'\right) \right| + \left|h_0\left(x;j\right)-h_0\left(x;j'\right) \right|
\Big)
\end{equation}
where $\tilde{p}$ is some intermediate value (in the whole argument) between $p_{1,0}\left(\,\cdot\,;j\right)$ and $p_{1,0}\left(\,\cdot\,;j'\right)$. We are in the hypothesis of Lemma \ref{lemF}, thus the underlying equality in \eqref{75} can be solved in $\left|m_1\left(x;j\right)-m_1\left(x;j'\right)\right|$, moreover:
\begin{equation}
\sup_{0<x\le\vei}\left|m_1\left(x;j\right)-m_1\left(x;j'\right)\right|\le c_{\delta_0} \left\|h_0\left(\,\cdot\,;j\right)-h_0\left(\,\cdot\,;j'\right)\right\|_{\infty}\le c_{\delta_0} \gamma_0' \left|j-j'\right|.
\end{equation}
Concerning the variation of the magnetic field $h_1$, we have:
\begin{equation}
\left|h_1\left(x;j\right)-h_1\left(x;j'\right)\right|\le 2\beta \gamma_0'^2\,\Big(\max\big\{\left|j\right|,\left|j'\right|\big\}\Big)\left\|m_1\left(\,\cdot\,;j\right)-m_1\left(\,\cdot\,;j'\right)\right\|_{\infty}.
\end{equation}
\qed
\end{comment}

\subsection{Lipschitz continuity for any $n$}

\begin{prop}
The sequence $\big(m_n\left(\,\cdot\,;j\right)\big)_{n\in\mathbb{N}}$ is Lipschitz continuous in $j$.
\end{prop}
\noindent
\textbf{Proof.} Suppose that for any $2\le n\le\bar{n}$, $\bar{n}$ fixed:
\begin{equation}
\left\| h_n\left(\,\cdot\,;j\right)-h_n\left(\,\cdot\,;j'\right)\right\|_{\alpha}\le \tau \left\| h_{n-1}\left(\,\cdot\,;j\right)-h_{n-1}\left(\,\cdot\,;j'\right)\right\|_{\alpha}
\end{equation}
with $\tau\in\left(0,1\right)$. Similarly to Lemma \ref{lemmaalpha}, it can be proved that this implies
\begin{equation}\label{uguale}
\left\| m_{n+1}\left(\,\cdot\,;j\right)-m_{n+1}\left(\,\cdot\,;j'\right)\right\|_{\alpha}\le r_0\left\| h_n\left(\,\cdot\,;j\right)-h_n\left(\,\cdot\,;j'\right)\right\|_{\alpha}.
\end{equation}
Thus, there exists a constant $c_{j,j'}>0$ such that for any $x\in(0,\vei]$ 
\begin{equation}
\left| h_{\bar{n}+1}\left(x;j\right)-h_{\bar{n}+1}\left(x;j'\right)\right|\le {c_{j,j'}\mathrm{e}^{\alpha\ve x}\over\alpha}\left\|m_{\bar{n}+1}\left(\,\cdot\,;j\right)-m_{\bar{n}+1}\left(\,\cdot\,;j'\right)\right\|_{\alpha}.
\end{equation}
Multiplying both members by $\mathrm{e}^{-\alpha\ve x}$ and taking the supremum with respect to $x$ we get
\begin{eqnarray}
\left\| h_{\bar{n}+1}\left(\,\cdot\,;j\right)-h_{\bar{n}+1}\left(\,\cdot\,;j'\right)\right\|_{\alpha}&\le& {c_{j,j'}\over\alpha}\left\|m_{\bar{n}+1}\left(\,\cdot\,;j\right)-m_{\bar{n}+1}\left(\,\cdot\,;j'\right)\right\|_{\alpha}\nn\\
&\le& {c_{j,j'}r_0\over\alpha}\left\| h_{\bar{n}}\left(\,\cdot\,;j\right)-h_{\bar{n}}\left(\,\cdot\,;j'\right)\right\|_{\alpha}
\end{eqnarray}
thus, if $\alpha>c_{j,j'}r_0$, $\tau\coloneqq c_{j,j'}r_0/\alpha$ is less than 1; furthermore
\begin{equation}
\left\| h_{\bar{n}}\left(\,\cdot\,;j\right)-h_{\bar{n}}\left(\,\cdot\,;j'\right)\right\|_{\alpha}\le {\tau\over 1-\tau}\left\| h_{0}\left(\,\cdot\,;j\right)-h_{0}\left(\,\cdot\,;j'\right)\right\|_{\alpha}\le {\tau\gamma'\over 1-\tau}\left|j-j'\right|
\end{equation}
and, by \eqref{uguale}:
\begin{equation}
\left\| m_{\bar{n}}\left(\,\cdot\,;j\right)-m_{\bar{n}}\left(\,\cdot\,;j'\right)\right\|_{\alpha}\le {\tau\gamma' r_0\over 1-\tau}\left|j-j'\right|.
\end{equation}
\qed

\begin{cor}\label{core}
The solution $m\left(\,\cdot\,;j\right)$ of \eqref{systm} obtained through the scheme established by Propositions \ref{p.3}, \ref{p.4} is Lipschitz continuous in $j$.
\end{cor}
\noindent
\textbf{Proof.} It is a straightforward consequence of uniform continuity of $\big(m_n\left(\,\cdot\,;j\right)\big)_{n\in\mathbb{N}}$.
\newline
\qed

\subsection{Proof of Theorem 2.2}

\noindent
In order to close the proof of Theorem 2.2 we need the following result:
\begin{prop}\label{prope}
For any $\mu\in\mathscr{M}_{\beta}$ there are $j^-\left(\mu\right)$ and $j^+\left(\mu\right)$ such that
\begin{equation}
m\,(\vei;j^-\left(\mu\right))=\nu^-, \qquad m\,(\vei;j^+\left(\mu\right))=\nu^+
\end{equation}
with $\nu^-<\nu<\nu^+$, where $\nu\coloneqq m\,(\vei;j\left(\mu\right))$.
\end{prop} 
\noindent
\textbf{Proof.} Indeed, Proposition \ref{p.5} is a combination of Corollary \ref{core} and Proposition \ref{prope}, that are enough to apply Intermediate Value Theorem. Define $\mu^-\coloneqq \mu-\eta$, $\mu^+\coloneqq \mu+\eta$, $\eta>0$ such that $\mu-\eta>m^*\left(\beta\right)$ and $\mu+\eta<1$. Recalling the definition of $j$, we define
\begin{equation}
j^{\pm}\left(\mu\right)=j\left(\mu\right) \mp \Big(\left(\chi_{\beta}\left(\mu\right)-1\right)\delta +\beta\mu\eta^2-{\beta\over 3}\eta^3\Big).
\end{equation}
Since we proved that there is $L>0$ so that $\left|\mu-\nu^{\pm}\right|<L\left|j-j^{\pm}\right|$ and that $\left|\mu-\nu\right|=O\left(\sqve\right)$, choose $\eta$ large enough and $\ve$ small enough to get the result.
\newline
\qed

\section*{Acknowledgements}
\noindent
We want to express our gratitude to professor Errico Presutti, who solved the problem in the free-boundary case and gave valuable hints to direct the rest of the proof. We also thank Matteo Colangeli, Anna De Masi, Cristian Giardin\`a and Dimitrios Tsagkarogiannis for enlightening discussions.


\vspace{-3mm}

\begin{thebibliography}{cccc}

\bibitem{GIAC}
G. B. Giacomin, J. L. Lebowitz, \textit{Phase segregation dynamics in particle system with long range interactions}, Journal of Statistical Physics 87(1) (1997): 37-61.
\bibitem{NERNST}
W. Nernst, Z. physik. Chem. 2 (1888): 613.
\bibitem{ONS}
L. Onsager, \textit{Theories and problems of liquid diffusion}, Annals of the New York Academy of Sciences 46(5) (1945): 241-265.
\bibitem{DARKEN1}
L. S. Darken, B. M. Larsen, \textit{Distribution of manganese and of sulphur between slag and metal in the open-hearth furnace}, Trans. Aime 150 (1942): 87-112.
\bibitem{DARKEN2}
L. S. Darken, \textit{Diffusion, mobility and their interrelation through free energy in binary metallic systems}, Trans. Aime 175 (1948): 184-201.
\bibitem{LSD}
L. S. Darken, \textit{Diffusion of carbon in austenite with a discontinuity in composition}, Trans. Aime 180(53) (1949): 430-438.
\bibitem{KR1}
R. Krishna. \textit{Uphill diffusion in multicomponent mixtures} Chemical Society Reviews 44(10) (2015): 2812-2836.
\bibitem{KR2}
R. Krishna. \textit{Serpentine diffusion trajectories and the Ouzo effect in partially miscible ternary liquid mixtures} Physical Chemistry Chemical Physics 17(41) (2015): 27428-27436.
\bibitem{EP1}
A. De Masi, E. Orlandi, E. Presutti, L. Triolo, \textit{Uniqueness and global stability of the instanton in nonlocal evolution equations} Rendiconti di Matematica e delle sue Applicazioni 14 (1994): 693-723.
\bibitem{EP2}
A. De Masi, E. Orlandi, E. Presutti, L. Triolo, \textit{Stability of the interface in a model of phase separation} Proceedings of the Royal Society of Edinburgh Section A: Mathematics 124(5) (1994): 1013-1022.
\bibitem{EP}
E. Presutti, \textit{Scaling limits in statistical mechanics and microstructures in continuum mechanics}.
Springer Science \& Business Media, 2008.
\bibitem{JL1}
L. Bertini, A. De Sole, D. Gabrielli, G. Jona-Lasinio, C. Landim, \textit{Fluctuations in stationary nonequilibrium states of irreversible processes}, Physical Review Letters 87(4) (2001): 040601.
\bibitem{JL2}
L. Bertini, A. De Sole, D. Gabrielli, G. Jona-Lasinio, C. Landim, \textit{Macroscopic fluctuation theory for stationary non-equilibrium states}, Journal of Statistical Physics, 107(3-4): 635-675.
\bibitem{JL3}
L. Bertini, A. De Sole, D. Gabrielli, G. Jona-Lasinio, C. Landim, \textit{Current fluctuations in stochastic lattice gases}, Physical Review Letters 94(3): 030601.
\bibitem{CDMP}
M. Colangeli, A. De Masi, E. Presutti, \textit{Particle models with self sustained current}, Journal of Statistical Physics 167(5) (2017): 1081-1111.
\bibitem{CGGV}
M. Colangeli, C. Giardin\`a, C. Giberti, C. Vernia, \textit{Nonequilibrium two-dimensional Ising model with stationary uphill diffusion}, Physical Review E 97(3) (2018): 030103.
\bibitem{DMOPc}
A. De Masi, E. Olivieri, E. Presutti, \textit{Critical droplet for a non local mean field equation}, Markov Processes and Related Fields 6 (2000): 439-472.
\bibitem{DMPT}
A. De Masi, E. Presutti, D. Tsagkarogiannis, \textit{Fourier law, phase transitions and the stationary Stefan problem}, Archive for rational mechanics and analysis 201(2) (2011): 681-725.
\bibitem{RB}
R. Boccagna, \textit{Fick's law in non-local evolution equations}, Journal of Mathematical Physics 59(5) (2018): 053508.
\bibitem{DMOP}
A. De Masi, E. Olivieri, E. Presutti, \textit{Spectral properties of integral operators in problems of interface dynamics and metastability}, Markov Process. Related Fields 4(2) (1998): 27-112.
\bibitem{CDGP}
G. Carinci, A. De Masi, C. Giardin\`a, E. Presutti, \textit{Free boundary problems in PDE's and Particle Systems}, Springer brief in Mathematical Physics 12 (2016).
\bibitem{LP}
J. L. Lebowitz, O. Penrose, \textit{Rigorous treatment of the Van der Waals-Maxwell theory of the liquid-vapor transition}, Journal of Mathematical Physics 7 (1966): 98-113.
\bibitem{DLSS}
B. Derrida, J. L. Lebowitz, E. R. Speer, H. Spohn, \textit{Fluctuations of a stationary nonequilibrium interface}, Physical Review Letters 67(2) (1991): 165.
	
\end{thebibliography}
\end{document}